\documentclass[conference, onecolumn, 11pt]{IEEEtran}

\IEEEoverridecommandlockouts
\IEEEpubid{\small \textcolor{blue}{Presented at the 59th Annual Allerton Conference on Communication, Control, and Computing, Monticello, IL, Sep., 2023.}}
\usepackage{cite}
\usepackage{amsmath,amssymb,amsfonts,amsthm,commath}
\usepackage{graphicx}
\usepackage{textcomp}
\def\BibTeX{{\rm B\kern-.05em{\sc i\kern-.025em b}\kern-.08em
    T\kern-.1667em\lower.7ex\hbox{E}\kern-.125emX}}

\usepackage{pifont}
\usepackage{multicol}
\usepackage{lettrine}
\usepackage{comment}
\usepackage{enumitem}
\usepackage{booktabs}
\usepackage[table]{xcolor}
\usepackage{bbm}
\usepackage{bm}
\usepackage{chngpage}
\usepackage{textcomp}
\usepackage{hyperref}
\usepackage{url}
\usepackage{orcidlink}
\usepackage[normalem]{ulem}
\usepackage{algpseudocode}
\newsavebox{\ieeealgbox}

\usepackage{array}

\newtheorem{theorem}{Theorem}
\newtheorem*{theorem*}{Theorem}

\newtheorem{lemma}{Lemma}
\newtheorem{definition}{Definition}

\newtheorem*{policy*}{OEs-aware Dynamic NEM}

 \def\old#1{}

\def\beq{\begin{equation}}
\def\eeq{\end{equation}}
\def\bea{\begin{eqnarray}}
\def\eea{\end{eqnarray}}
\def\ba{\begin{array}}
\def\ea{\end{array}}

\def\bitem{\begin{itemize}}
\def\eitem{\end{itemize}}
\def\ben{\begin{enumerate}}
\def\een{\end{enumerate}}

\definecolor{bgrd}{rgb}{1,1,1}
\definecolor{gray}{rgb}{0.5,0.5,0.5}
\definecolor{dkr}{rgb}{0.7,0.1,0.2}
\definecolor{dkb}{rgb}{0.1,0.1,0.8}

\def\tcb{\textcolor{blue}}

\begin{document}

\title{Operating-Envelopes-Aware Decentralized Welfare Maximization for Energy Communities
}
% OLD title
%Decentralized Optimization for Energy Communities under Operating Envelopes
\author{\IEEEauthorblockN{Ahmed S. Alahmed\orcidlink{0000-0002-4715-4379}, Guido Cavraro\orcidlink{0000-0003-0296-720X}, Andrey Bernstein\orcidlink{0000-0003-4489-8388}, and Lang Tong\orcidlink{0000-0003-3322-2681}\thanks{\scriptsize Ahmed S. Alahmed and Lang Tong are with the School of Electrical and Computer Engineering, \textit{Cornell University}, Ithaca, NY, USA ({\tt \{\tcb{ASA278,~LT35}\}\tcb{@cornell.edu}}). Guido Cavraro and Andrey Bernstein are with the Power System Engineering Center, \textit{National Renewable Energy Laboratory}, Golden, CO, USA ({\tt \{\tcb{GCAVRARO,~ABERNSTE}\}\tcb{@nrel.gov}}).
}
}}

\maketitle

\IEEEpubidadjcol
\begin{abstract}
We propose an operating-envelope-aware, prosumer-centric, and efficient energy community that aggregates individual and shared community distributed energy resources and transacts with a regulated distribution system operator (DSO) under a generalized net energy metering tariff design. To ensure safe network operation, the DSO imposes dynamic export and import limits, known as dynamic operating envelopes, on end-users' revenue meters. Given the operating envelopes, we propose an incentive-aligned community pricing mechanism under which the decentralized optimization of community members' benefit implies the optimization of overall community welfare. The proposed pricing mechanism satisfies the cost-causation principle and ensures the stability of the energy community in a coalition game setting. Numerical examples provide insights into the characteristics of the proposed pricing mechanism and quantitative measures of its performance.
\end{abstract}


\section{Introduction}\label{sec:intro}
\lettrine{D}{espite the} ambitious electrical grid decarbonization goals by increasing the penetration of behind-the-meter (BTM) distributed energy resources (DER), many end-users are ineligible to install BTM DER due to several physical, financial, and jurisdictional challenges\footnote{The National Renewable Energy Laboratory (NREL) reported that $\sim 75\%$ of households in the U.S. are ineligible for rooftop solar installations \cite{NREL_CommunitySolar:08NREL}.}. {\em Energy communities} overcome many DER adoption hurdles by allowing a group of spatially co-located customers to pool and aggregate their resources and perform energy and monetary transactions as a single entity behind the DSO's revenue meter \cite{Yang&Guoqiang&Spanos:21TSG}. Under the widely adopted net energy metering (NEM) policy design \cite{NEMevolution:23NAS,Alahmed&Tong:22EIRACM}, the meter measures the community's net consumption and assigns a {\em buy (retail) rate} if the community is net-importing, and a {\em sell (export) rate} if the community is net-exporting \cite{Alahmed&Tong:22EIRACM}. Enabling jurisdictions and programs, such as NEM aggregation (NEMA) and virtual NEM (VNEM), play a critical role in the proliferation of energy communities \cite{NEMevolution:23NAS}.

Without proper coordination of the immense flexibility that DER introduces, DER imports and exports can result in two-way energy flows that threaten the voltage and thermal limits of the distribution networks \cite{Liu&Ochoa&Riaz&Mancarella&Ting&San&Theunissen:21PEM}. The high imports and exports issue is exacerbated by the fact that, to the DSO, such resources are neither visible (due to load masking) nor controllable (due to the unbundled model of deregulated electricity markets). Therefore, to ensure that the community's power exports and imports do not compromise the distribution network's operation, the DSO may impose dynamic operating envelopes (OEs\footnote{Although the acronym "DOE" is more widely used to refer to dynamic OEs, we avoid it here, as it is usually used to acronymize the U.S. Department of Energy.}) on its end-users revenue meters. The OEs, which may vary temporally and spatially depending on the network's conditions, provide much higher flexibility over the widely adopted scheme {\em fixed export limits} (e.g., 5kW or 3.5kW \cite{Liu&Ochoa&Riaz&Mancarella&Ting&San&Theunissen:21PEM}), which quickly become obsolete as the BTM DER penetration level grows.

In this work, we propose a network-aware energy community market mechanism that induces its members to maximize global welfare. The mechanism's OEs-aware, resource-aware, and threshold-based pricing and payment rules ensure that the community's aggregate flexible demand schedule is actively adapting to its aggregate supply, which in turn maximizes the community's welfare.

Despite the abundant previous work on energy communities, the majority of literature that considered pricing-based energy management market mechanisms and cost allocation rules neglected network and grid constraints \cite{Yang&Guoqiang&Spanos:21TSG,Fleischhacker&Corinaldesi&Lettner&Auer&Botterud:22TSG,Han&Morstyn&McCulloch:19TPS,Chen&Zhao&Low&Mei:21TSG,Chakraborty&Poolla&Varaiya:19TSG,Elrahi&Etesami&Saad&Mandayam&Poor:19TSG, Vespermann&Hamacher&Kazempour:21TPS}. On the other hand, the literature on OEs largely ignored incorporating them into a price-driven mechanism design that induces community members to collectively react to ensure a safe community operation \cite{Liu&Ochoa&Wong&Theunissen:21TSG,Blackhall:20ARENA,Gerdroodbari&Khorasany&Razzaghi:22AE,Yi&Verbic:22EPSR}. The study in \cite{Azimetal:23TSG}, which considers an energy community with operator-designed OEs to maximize energy transactions among its members without compromising network constraints, is perhaps the closest to our work. However, the authors adopt an {\em ex-post} allocation rule, namely Shapley value, to distribute the coalition welfare, whereas in our case, {\em ex-ante} and resource-aware pricing and allocation mechanisms are designed to distribute the coalition welfare and incentivize joining the coalition.

To the best of our knowledge, no previous work incorporated OEs into a price-driven and welfare-maximizing market mechanism design that induces rational prosumers to join the community while meeting the {\em cost-causation principle}.

To this end, we propose an OEs-aware and welfare-maximizing market mechanism for energy communities that aggregates individual and shared community resources under a general NEM policy. The proposed market mechanism 
\begin{itemize} 
\item[\ding{72}] incorporates the DSO-imposed OEs at the end-user level, ensuring that the community's operation is network-aware. 
\item[\ding{72}] guarantees surplus levels to its members higher than the maximum attainable surplus under standalone settings.
\item[\ding{72}] decentrally achieves welfare optimality. 
\item[\ding{72}] satisfies the generalized {\em cost-causation principle}. 
\end{itemize}

Under the proposed market mechanism, the community operator charges/compensates its members via a two-threshold-based dynamic price that is a function of the DERs in the community. The community price is uniform to all members, regardless of the homogeneity of the OEs, and whether the OEs are binding or not. The price monotonically decreases as the community aggregate generation/load ratio increases, indicating that the excess generation from net-producing members is first pooled with net-consuming members before it is exported back to the grid.

This paper generalizes our {\em Dynamic NEM} (D-NEM) mechanism \cite{Alahmed&Tong:23ECjournalarXiv, Alahmed&Tong:23ECPESGM} in three different aspects\footnote{For the rest of this paper, we, interchangeably, use {\em OEs-aware D-NEM} and {\em D-NEM} to refer to the proposed market mechanism in this work.}. First, incorporating OEs gives rise to a community price with a threshold structure that requires knowledge of the BTM generation and OEs of every member. Second, unlike D-NEM, the community member's optimal decision is also a threshold policy that is computed using the announced community price and the assigned OEs. Third, the decision problem of the OEs-aware benchmark customer under the DSO is a four-threshold policy.

In the next section, we introduce the OEs-constrained energy community framework and the available DER, in addition to the benchmark model outside the community. In Section \ref{sec:MktMech}, we present the OE-aware D-NEM mechanism and the community member decision problem. In Section \ref{sec:theory}, we present the decentralized welfare optimality and cost-causation conformity results under the proposed OEs-aware D-NEM, followed by a numerical study to showcase the community performance compared to the benchmark in Section \ref{sec:num}, and a summary of our findings in Section \ref{sec:conclusion}. %extending the efficiency, individual rationality, and equity theorems to import/export-constrained communities. 

\section{Energy Sharing Model and Benchmark}\label{sec:model}
Here, we describe the energy community structure, resources, and  payment and surplus functions in sections \ref{subsec:structure}-\ref{subsec:surplusWelfare}, followed by establishing the standalone-DSO-prosumer benchmark model in section \ref{subsec:benchamrk}, which is the reference model for community members. Lastly, we present the cost-causation principle in section \ref{subsec:costCausation}.

\subsection{Energy Community Structure}\label{subsec:structure}
\begin{figure}[htbp]
    \centering
    \includegraphics[scale=0.52]{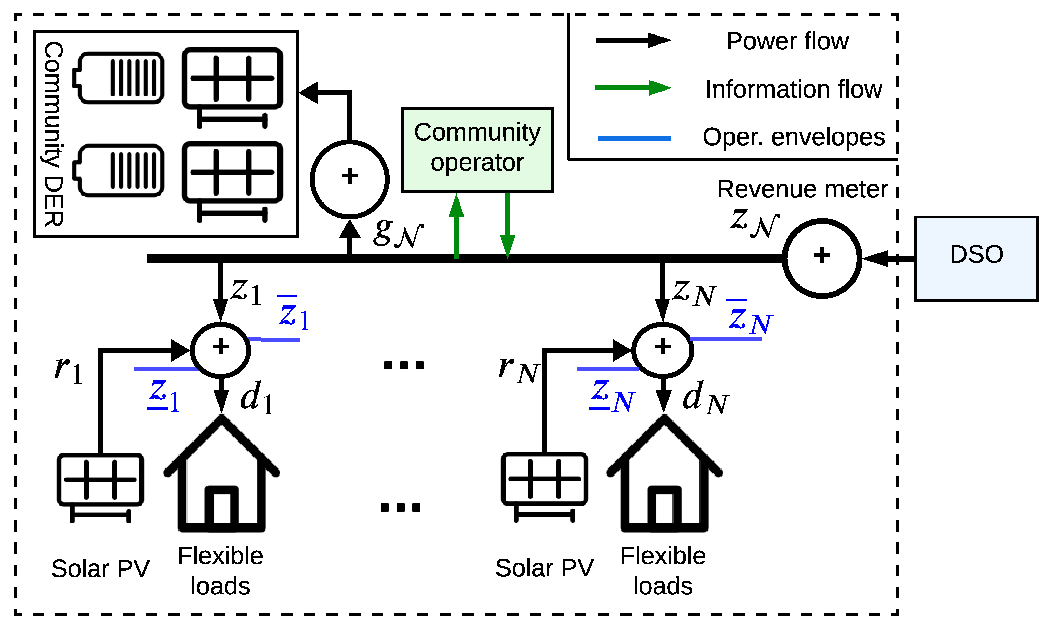}
    \caption{Energy community framework. Member consumption and renewables are $d_i \in \mathbb{R}_+, r_i \in \mathbb{R}_+$, respectively, and member and aggregate net consumption, and community (central) DER are $z_i\in \mathbb{R},z_\mathcal{N} \in \mathbb{R}, g_\mathcal{N}\in \mathbb{R}_+$, respectively. The direction of the arrows indicates positive quantities.}
    \label{fig:EnergyCommunity}
\end{figure}

As a single entity behind the DSO's NEM revenue meter (i.e., point of common coupling (PCC)), the profit-neutral operator receives one bill on behalf of its $N$ community members, represented by the set $\mathcal{N}:=\{1,\ldots,N\}$, who are subject to the operator's market mechanism that determines the pricing and payment rules (Fig.\ref{fig:EnergyCommunity}). Community network constraints are incorporated by considering the DSO-imposed OEs at each member's revenue meter, which guarantee the safe operation of the community. The operator's goal is to devise a market mechanism that incorporates the DSO-imposed OEs and announces a price that induces its members to achieve social welfare optimality in a decentralized fashion (Fig.\ref{fig:EnergyCommunity}). Given the market mechanism, each community member optimizes its own resources subject to the DSO-imposed OEs and other consumption constraints. 

Before we model the local (i.e., BTM) and community (central) resources, we assume that the BTM renewable distributed generation (DG) outputs of every member are available to the operator through sub-meters. With little loss of generality, we assume that the DSO's NEM {\em netting frequency} \cite{Alahmed&Tong:22EIRACM} is commensurate with the frequency at which it announces the OEs.

\subsection{Energy Community DER}\label{subsec:resources}
The community resources are either located within prosumer premises, i.e., behind their revenue meters; to which we refer as {\em BTM DER}, or they are located in front of their meters but still downstream of the PCC; to which we refer to as {\em community DER}. 
\subsubsection{BTM and community shared DER}\label{subsec:ModelBTM}
Each community member may have flexible loads and solar PV. We assume each member $i\in \mathcal{N}$ is equipped with $K$ devices, represented by the set $\mathcal{K}:=\{1,\ldots,K\}$ whose {\em load consumption} is denoted by
\begin{equation}\label{eq:Conslimit}
\bm{d}_i = (d_{i1},\cdots, d_{iK}) \in \mathcal{D}_i:=\{\bm{d}_i:\underline{\bm{d}}_i \preceq \bm{d}_i \preceq \overline{\bm{d}}_i\} \subseteq \mathbb{R}^K_+,
\end{equation}
where $ \underline{\bm{d}}_i$ and $\overline{\bm{d}}_i$ are the device bundle’s lower and upper consumption limits of $i\in \mathcal{N}$, respectively. For a critical inflexible load $k \in \mathcal{K}$, prosumer $i$ sets $\underline{d}_{ik} = \overline{d}_{ik}$. The community {\em aggregate consumption} is defined as $d_\mathcal{N}:= \sum_{i \in \mathcal{N}} \bm{1}^\top \bm{d}_i$.

We assume that each member may have access to private BTM PV output and a share of community PV production\footnote{Here, we ignore community and/or BTM storage. See \cite{Alahmed&Tong:23ECjournalarXiv} for storage incorporation.}.  Let $b_i\in \mathbb{R}_+$ be the renewable generation of member $i$ from both BTM generation and community generation. Therefore, the aggregate generation of all members, including community generation, is given by $b_\mathcal{N}:=\sum_{i \in \mathcal{N}} b_i$. The vector of renewable DG of members is denoted by $\bm{b}:=(b_1,\ldots,b_N)\in \mathbb{R}^N_+$.
\subsubsection{Net consumption}\label{subsec:ModelNetCons}
Given the outputs of the BTM DER and share of the community DER, the {\em net consumption} of every member $i\in \mathcal{N}$ is defined as 
\begin{equation}\label{eq:Netconsi}
    z_i:= \bm{1}^\top \bm{d}_i - b_i \in \mathcal{Z}_i:=\{ z_i:\underline{z}_i\leq z_i\leq \overline{z}_i],
\end{equation}
where $z_i>0$ ($z_i<0$) indicates a {\em net-consuming} ({\em net-producing}) individual, and $\underline{z}_i \leq 0$ and $\overline{z}_i \geq 0$ are the DSO-imposed export and import OEs at the prosumer's revenue meter, respectively\footnote{Without loss of generality, the OEs $\underline{z}_i,\overline{z}_i$ are adjusted, for every $i$, to be cognizant of the virtual generation share $g_i$.}. The community {\em aggregate net consumption} is given by
\begin{equation}\label{eq:NetconsN}
    z_\mathcal{N}:= \sum_{i\in \mathcal{N}} z_i = d_\mathcal{N} - b_\mathcal{N},
\end{equation}
where $ z_\mathcal{N}>0$ ($z_\mathcal{N}<0$) indicates a {\em net-consuming} ({\em net-producing}) community.

\subsection{Energy Community Payments and Surpluses}\label{subsec:surplusWelfare}
\subsubsection{Community Payments and Profit}
At the PCC, the community faces the DSO's NEM X tariff model \cite{Alahmed&Tong:22IEEETSG}, characterized by the parameter $\pi^{\mbox{\tiny NEM}}=(\pi^+,\pi^-)$, which has a {\em pricing rule} $\Gamma^{\mbox{\tiny NEM}}$ and a {\em payment rule} $P^{\mbox{\tiny NEM}}_\mathcal{N}$
 \begin{align}
       \Gamma^{\mbox{\tiny NEM}}(z_\mathcal{N}) = \begin{cases}
\pi^+, &\hspace{-0.85em} z_\mathcal{N}\geq 0 \\ 
\pi^-, &\hspace{-0.85em} z_\mathcal{N}< 0
\end{cases},~
    P^{\mbox{\tiny NEM}}_\mathcal{N}(z_\mathcal{N}) = \Gamma^{\mbox{\tiny NEM}}(z_\mathcal{N}) \cdot z_\mathcal{N},\label{eq:Pcommunity}
    \end{align}
respectively, where $\pi^+\geq 0$ and $\pi^- \geq 0$ are the {\em buy} (retail) and {\em sell} (export) rates\footnote{Here we do not include possible fixed charge in NEM tariff $\pi^0$, assuming that such a fixed charge is matched by membership fees.}. We assume that $\pi^+ \geq \pi^-$, which avoids risk-free price arbitrage, given that the retail and export rates are deterministic and known apriori.

The role of the community operator is to come up with a {\em community pricing rule} $\chi: \bm{b} \rightarrow \Gamma^\chi(\cdot)$ for its members that incentivizes the members toward achieving the maximum social welfare. Given the pricing rule $\chi$, we define the vector of payment (allocation) of community members as functions of individual net consumption by $\bm{P}_i^\chi(\bm{z}):=(P_1^\chi(z_i),\ldots,P_N^\chi(z_N))$, where $\bm{z}:=(z_1,\ldots,z_N)\in \mathbb{R}^N$. 

To achieve profit neutrality, the community operator must ensure that the money it pays to the utility matches the aggregated payments of its members, i.e.,
\begin{equation}\label{eq:ProfitNeutrality}
    \sum_{i\in \mathcal{N}} P^\chi_i(z_i) - P^{\mbox{\tiny NEM}}_\mathcal{N}(z_\mathcal{N}) = 0.
\end{equation} 
%where $\bm{z}:=(z_1,\ldots,z_N)\in \mathbb{R}^N$.

\subsubsection{Community Members Surplus and Decision Problem}
The {\em surplus} of every $i \in \mathcal{N}$ community member is characterized by comfort/satisfaction and economics metrics as
\begin{equation}\label{eq:Surplusi}
    S^{\chi}_i(\bm{d}_i,z_i):= \underbrace{U_i(\bm{d}_i)}_\text{utility of consumption}-\underbrace{P^{\chi}_i(z_i)}_\text{payment under $\chi$},
\end{equation}
where, for every member $i\in \mathcal{N}$, the {\em utility of consumption function} $U_i(\bm{d}_i)$ \cite{Varian2microeconomic:1992book} is assumed to be additive, concave, non-negative, non-decreasing, and continuously differentiable with a {\em marginal utility function} $\mathbf{L}_i:=\nabla U_i=\left(L_{i 1}, \ldots, L_{i K}\right)$. For notational simplicity, we denote the {\em inverse marginal utility} vector by $\bm{f}_i:=(f_{i1},\ldots,f_{iK})$, where $f_{ik}:= L^{-1}_{ik}, \forall i\in \mathcal{N}, \forall k\in \mathcal{K}$. Note that adopting the {\em surplus} function to characterize the prosumer's {\em payoff} (benefit) is more general than using the {\em payment} function. 

After announcing the pricing rule, every community member $i \in \mathcal{N}$ solves the following surplus-maximization program
 \begin{align} \label{eq:SurplusMemberOpt}
\mathcal{P}_{i}^{\chi}:&~ \underset{\bm{d}_i \in \mathbb{R}_{+}^K, z_i \in \mathbb{R}}{\operatorname{maximize}} ~~~ S_i^{\chi}(\bm{d}_i,z_i):=U_i\left(\bm{d}_i\right)- P^{\chi}_i(z_i) \nonumber \\
& \text { subject to } \quad z_i := \boldsymbol{1}^\top \bm{d}_{i}-b_i \\
& \hspace{2.4cm}
\underline{\bm{d}}_i \preceq \bm{d}_i \preceq \overline{\bm{d}}_i\nonumber \\
& \hspace{2.4cm} \underline{z}_i \leq z_i \leq \overline{z}_i. \nonumber 
\end{align} 
Denote the optimal value function of (\ref{eq:SurplusMemberOpt}) by $S_i^{\ast,\chi}(b_i) := S_i^{\ast,\chi}(\bm{d}^{\ast,\chi}(b_i),z^{\ast,\chi}(b_i))$. To ensure that a feasible solution to (\ref{eq:BenchmarkProblem}) always exists, we assume that, for every $i \in \mathcal{N}$, the OEs ($\overline{z}_i,\underline{z}_i$) satisfy 
 \begin{equation}\label{eq:feasibilityAssum}
 \overline{z}_i\geq \bm{1}^\top \underline{\bm{d}}_{i} - b_i,~~~~~  \underline{z}_i\leq \bm{1}^\top \overline{\bm{d}}_{i} - b_i.
 \end{equation}

\subsection{Benchmark Model, Surplus and Decision Problem}\label{subsec:benchamrk}
To ensure fairness when comparing the surplus of community members under the proposed market mechanism to their benchmark surplus, i.e., under the DSO's regime, we posit that prosumer resources (BTM DER, and share from central generation) are the same with and without the community. 

The {\em benchmark payment}, for every member $i\in \mathcal{N}$, is given by the DSO's NEM X tariff as $P^{\mbox{\tiny NEM}}(z_i)$.

Similar to community members, the {\em surplus} of the {\em benchmark} prosumer, for every $i\in \mathcal{N}$, is defined as
\begin{equation}\label{eq:SurplusBench}
S^{\mbox{\tiny NEM}}_i(\bm{d}_i, z_i):= U_i(\bm{d}_i)-P^{\mbox{\tiny NEM}}(z_i).
\end{equation}
Each surplus-maximizing benchmark prosumer solves
\begin{align} \label{eq:BenchmarkProblem}
\mathcal{P}_{i}^{\mbox{\tiny NEM}}:   \underset{\bm{d}_i \in \mathbb{R}^K_+, z_i \in \mathbb{R}}{\rm maximize}~~&  S^{\mbox{\tiny NEM}}_i(\bm{d}_i, z_i):=U_i(\bm{d}_i) - P^{\mbox{\tiny NEM}}(z_i) \nonumber \\\text{subject to}~~ & z_i := \boldsymbol{1}^\top \bm{d}_{i} - b_i \\&
\underline{\bm{d}}_i \preceq \bm{d}_i \preceq \overline{\bm{d}}_i \nonumber\\& \underline{z}_i \leq  z_i \leq \overline{z}_i. \nonumber
\end{align} 
 Lemma \ref{lem:BenchmarkSur} in the appendix characterizes the benchmark's maximum surplus function $$S^{\ast,\mbox{\tiny NEM}}_i(b_i):=S^{\ast,\mbox{\tiny NEM}}_i(\bm{d}^{\ast,\mbox{\tiny NEM}}_i(b_i),z^{\ast,\mbox{\tiny NEM}}_i(b_i)),$$ and shows that it is a monotonically increasing function of prosumer's renewables $b_i$.

 \subsection{Generalized Cost-Causation Principle}\label{subsec:costCausation}
To ensure that the pricing rule $\chi$ is just we use the {\em cost-causation principle} developed in \cite{Chakraborty&Poolla&Varaiya:19TSG}, but with a generalization that incorporates surplus-based individual rationality rather than payment-based one.
\begin{definition}[Generalized cost-causation principle]\label{def:CostCausation}
    A market mechanism that achieves the following five axioms is a cost-causation-based market mechanism.
    \begin{description}[style=unboxed,leftmargin=0cm]
  \item[Axiom 1] (Individual rationality). The surplus of every $i\in \mathcal{N}$ community member should be at least equal to its benchmark surplus, i.e., $S_i^{\ast,\chi} \geq S^{\ast,\mbox{\tiny NEM}}_i, \forall i \in \mathcal{N}$.
  \item[Axiom 2] (Profit neutrality). The market operator must be profit-neutral, i.e., (\ref{eq:ProfitNeutrality}) is satisfied.
  \item[Axiom 3] (Equal treatment of equals). The mechanism equally treats the equals, if, for any two community members $i,j\in \mathcal{N}, i\neq j$, having $z_i =z_j$ yields $P^{\chi}_i(z_i) = P^{\chi}_j(z_j)$.
  \item[Axiom 4] (Monotonicity). The mechanism is monotonic if, for any two community members $i,j\in \mathcal{N}, i\neq j$, having $\abs{z_i}\geq \abs{z_j}$ and $z_i z_j \geq 0$ yields $|P^{\chi}_i(z_i)| \geq |P^{\chi}_j(z_j)|$.
  \item[Axiom 5] (Cost causation penalty and cost mitigation reward). A net-consuming (net-producing) community member $z_i>0$ ($z_i<0$) causes (mitigates) cost to (from) the community, and therefore should be penalized (rewarded) for it, i.e., for any $i\in \mathcal{N}$, if $z_i>0$ then $P^{\chi}_i(z_i)>0$, whereas if $z_j<0$ then $P^{\chi}_j(z_j)<0$.
\end{description}
\end{definition}

\section{Operating-Envelopes-Aware Dynamic NEM}\label{sec:MktMech}
The goal of the profit-neutral community operator is to devise an OE-aware market mechanism that achieves maximum welfare in a distributed fashion while satisfying the DSO OEs and the generalized cost-causation principle. 

\begin{definition}[Decentralized welfare optimality]
    The community welfare is decentrally maximized if there exists a pricing rule $\chi^\sharp$ such that the maximum welfare under centralized community operation
    \begin{align}\label{eq:CentralWelfare}
\mathcal{P}_{\mathcal{N}}^{\mbox{\tiny NEM}}:\underset{\{\bm{d}_i\}_{i=1}^N,\{z_i\}_{i=1}^N}{\rm maximize}& W_\mathcal{N}^{\mbox{\tiny NEM}}(\bm{d}_1,\ldots,\bm{d}_N,\bm{z}) := \mathbb{E} \bigg[\sum_{i=1}^N (U_{i}(\bm{d}_i) - P^{\mbox{\tiny NEM}}_\mathcal{N}(\sum_{i=1}^{N} z_i) )\bigg]\nonumber\\ \text{subject to} &~~~~ (\ref{eq:Conslimit})-(\ref{eq:Netconsi}), ~ \forall i\in \mathcal{N}\\&~~~~ (\ref{eq:NetconsN})-(\ref{eq:ProfitNeutrality}), \nonumber
\end{align} 
%% use this to make the objective two lines
%\nonumber\\&\hspace{3.5cm}
denoted by $W_\mathcal{N}^{\dagger,\mbox{\tiny NEM}}$, is achieved by the aggregate maximum surpluses of community members under the pricing rule $\chi^\sharp$, i.e., if
   \begin{equation}
       W_\mathcal{N}^{\dagger,\mbox{\tiny NEM}}(\bm{b}) = \sum_{i\in \mathcal{N}} S_{i}^{\ast,\chi^\sharp}(b_i).
   \end{equation}
\end{definition}
Heed that the assumption in (\ref{eq:feasibilityAssum}) guarantees the existence of a feasible solution to $\mathcal{P}_{\mathcal{N}}^{\mbox{\tiny NEM}}$ in (\ref{eq:CentralWelfare}).

\subsection{Operating-Envelopes-Aware Dynamic NEM}\label{subsec:MktMech}
Here, we propose the OEs-aware D-NEM, and show that the community price is announced without compromising members' privacy. Only the BTM and community-level renewable generation are needed to determine the community price.

\begin{comment}
By observing historical demand bids of its members (inverse marginal utilities $f_{ik}, \forall i \in \mathcal{N}, k\in \mathcal{K}$), which reflect their willingness to consume, the community operator estimates the inverse marginal utilities' parameters and functional forms\footnote{There is a plethora of regression and classification algorithms in machine learning that can be used for parameter estimation.}. Note that the privacy of community members is preserved, as their utility functions are unknown to the operator. Given the demand bids, OEs, and community generation-storage outputs $\bm{b}$, the operator envisages the following OEs-aware D-NEM.
\end{comment}

\begin{policy*}
The threshold-based, OEs-aware, community pricing $\Gamma^{\mbox{\tiny DNEM}}(\bm{b})$ and payment rules $P^{\mbox{\tiny DNEM}}_i$ are, respectively, given by the 3-tuple tariff parameter $\pi^{\mbox{\tiny DNEM}}=(\pi^+, \pi^z(\bm{b}), \pi^-)$ with the order $\pi^+ \geq \pi^z(\bm{b}) \geq \pi^-$, as
\begin{align}\label{eq:PricingMechanism}
 \Gamma^{\mbox{\tiny DNEM}}(\bm{b}) &= \begin{cases}
 \pi^+ & ,     b_\mathcal{N}< \sigma_1(\bm{b})  \\ 
\pi^z(\bm{b}) & , b_\mathcal{N}\in [\sigma_1(\bm{b}),\sigma_2(\bm{b})]\\ 
\pi^- & , b_\mathcal{N} > \sigma_2(\bm{b}),
\end{cases}\\
    P^{\mbox{\tiny DNEM}}_i (z_i) &= \Gamma^{\mbox{\tiny DNEM}} \cdot z_i, \qquad \forall i \in \mathcal{N}. \label{eq:PaymentRule}
 \end{align}
 where the thresholds $\sigma_1(\bm{b})$ and $\sigma_2(\bm{b})$ are computed as
$$
\begin{aligned}
\sigma_1(\boldsymbol{b})  :=\sum_{i=1}^N \max \left\{\underline{z}_i+b_i, \min \left\{R_i^{+}, \bar{z}_i+b_i\right\}\right\},~~
\sigma_2(\boldsymbol{b})  :=\sum_{i=1}^N \max \left\{\underline{z}_i+b_i, \min \left\{R_i^{-}, \bar{z}_i+b_i\right\}\right\} \geq \sigma_1(\boldsymbol{b}),
\end{aligned}
$$
and
$$
\begin{aligned}
 R_i^{+}:=\mathbf{1}^{\top} \max \left\{\underline{\boldsymbol{d}}_i, \min \left\{\boldsymbol{f}_i\left(\mathbf{1} \pi^{+}\right), \overline{\boldsymbol{d}}_i\right\}\right\},~~
 R_i^{-}:=\mathbf{1}^{\top} \max \left\{\underline{\boldsymbol{d}}_i, \min \left\{\boldsymbol{f}_i\left(\mathbf{1} \pi^{-}\right), \overline{\boldsymbol{d}}_i\right\}\right\},
\end{aligned}
$$
where the $\max$ and $\min$ operators are elementwise.

The price $\pi^z(\bm{b}):=\mu^\ast(\bm{b})\in (\pi^-,\pi^+)$ is the solution of
\begin{equation}\label{eq:MonotonicPrice}
    \sum_{i=1}^N \max \left\{\underline{z}_i+b_i, \min \left\{R_i^z(\mu), \bar{z}_i+b_i\right\}\right\}=b_{\mathcal{N}},
\end{equation}
where
$$
R_i^z(\mu):=\mathbf{1}^{\top} \max \left\{\underline{\boldsymbol{d}}_i, \min \left\{\boldsymbol{f}_i(\mathbf{1} \mu), \overline{\boldsymbol{d}}_i\right\}\right\} .
$$
\end{policy*}

\subsection{Operating-Envelopes-Aware Dynamic NEM Properties and Structure}
\begin{figure}
    \centering
    \includegraphics[scale = 0.4]{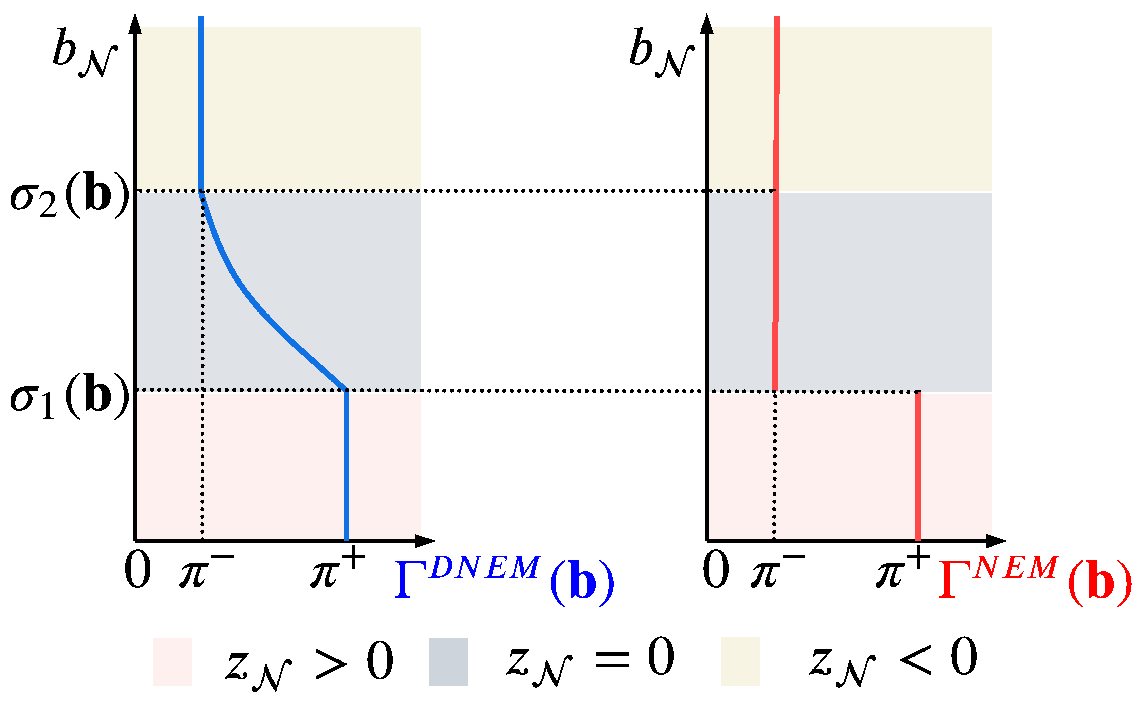}
    \caption{OEs-aware D-NEM and NEM prices under optimal community member response.}
    \label{fig:OE_DNEMprice}
\end{figure}

The OEs-aware D-NEM, shown in Fig.\ref{fig:OE_DNEMprice}, offers nice and intuitive structural properties. 
\begin{enumerate}[wide, labelindent=0pt]
    \item Resource- and OEs-aware pricing: The community price is a function of the centralized and BTM resources in the community. It also takes into account, the network constraints represented by the DSO-imposed OEs at every member's revenue meter.
    \item Threshold-based structure: The operator announces the prices based on a 2-thresholds ($\sigma_1,\sigma_2$) policy that partitions the range of $b_\mathcal{N}$ into three regions. The thresholds are computed in closed-from given the DSO's tariff and OEs, and the vector of measured DER $\bm{b}$ and prosumer bids.
    \item Privacy-preserving mechanism: The two thresholds ($\sigma_1,\sigma_2$) can be computed without compromising member's privacy. In particular, the values $R_i^+$ and $R_i^-$ are provided {\em apriori} by the community members, given the public utility prices. Also, given that $\pi^z(\bm{b})\in (\pi^-,\pi^+)$, the members can provide a value for each price sample, which are then used by the operator to compute the price $\pi^z(\bm{b})$ depending on the aggregate generation $b_\mathcal{N}$.
    \item Non-discriminatory pricing: The proposed network-aware community price is {\em uniform} to all members, even when some OEs may be binding, and regardless of the heterogeneity of the DSO-imposed OEs\footnote{Imposing OEs at the community's PCC rather than at the prosumers' revenue meters in our case might yield some form of discrimination, perhaps through fixed non-uniform re-allocations (analogous to uplifts in wholesale markets). A full analysis of the PCC OEs scheme is pursued in \cite{AllertonExtension:23}.}. 
    \item Supply/Demand balance: The community price dynamically decreases as the supply-to-demand ratio increases, which is economically intuitive. The decreasing (increasing) price as the supply-to-demand ratio increases (decreases) induces the community members to increase (decrease) their consumption, which reduces the community's net exports and imports, and brings it closer to energy balancedness. 
    \item Endogenously-determined market roles: Unlike conventional electricity markets, where the roles of \textit{buyers} and \textit{sellers} are predetermined, the proposed mechanism allows community members to determine their roles (see Section \ref{subsec:MemberProblem}).
    \item Scalability and explainability: The proposed pricing mechanism is tractable and highly scalable, as it scales linearly with the number of community members $\mathcal{O}(N)$. This is not the case under computationally intensive allocation rules, such as Nucleolus \cite{Han&Morstyn&McCulloch:19TPS,Schotter&Schwodiauer:80JEL}, Shapley value \cite{Shapley&Shubik:73RAND, Azimetal:23TSG}, and Nash bargaining \cite{Compte&Jehiel:10Econometrica}. Furthermore, the payment amount of each community member $i$ can be easily understood and justified \cite{Tsaousoglou&Giraldo&Paterakis:22RSER}, as it depends on their own net consumption $z_i$, which is not the case under, for example, Shapley value, that allocates payments to community members based on their marginal contributions to the coalition, which can only be computed by the market operator and may not be proportionate to members' net consumptions.
\end{enumerate}
The operator sets the price by comparing the aggregate DG output in the community $b_\mathcal{N}$ to the thresholds ($\sigma_1,\sigma_2$) that characterize the total willingness of prosumers to consume while factoring in the remaining headroom to reach their import/export limits (Fig.\ref{fig:OE_DNEMprice}). Fig.\ref{fig:OE_DNEMprice} depicts the community operator price $\Gamma^{\mbox{\tiny DNEM}} (\bm{b})$ (blue) and the DSO's NEM X price the community faces at the PCC $\Gamma^{\mbox{\tiny NEM}}(\bm{b})$, and shows that the thresholds partition the range of $b_\mathcal{N}$ into three zones based on the aggregate net consumption under members optimal decisions characterized in Lemma \ref{lem:OptSchedule}. 

When the OEs of all members are relaxed, i.e., $\overline{z}_i \rightarrow \infty, \underline{z}_i \rightarrow -\infty, \forall i \in \mathcal{N}$, the market mechanism converges to the one in \cite{Alahmed&Tong:23ECjournalarXiv}, and the thresholds ($\sigma_1, \sigma_2$) become independent of the renewables. More precisely, $\sigma_1(\bm{b}) \rightarrow R^+_i$ and $ \sigma_2(\bm{b}) \rightarrow R^-_i$. Conversely, tightening any of the import (export) envelopes $\overline{z}_i$ ($\underline{z}_i$) shifts both thresholds to the left (right), which reflects the effect of individual OEs on the community's aggregate net-consumption at the PCC $z_\mathcal{N}$, and therefore the community price $\Gamma^{\mbox{\tiny DNEM}}(\bm{b})$.

\subsection{Community Member Problem and Optimal Decisions}\label{subsec:MemberProblem}
 Given $b_\mathcal{N}$, the community pricing and payment rules are announced, and accordingly, every member $i \in \mathcal{N}$ solves the (\ref{eq:SurplusMemberOpt}), which we reformulate to:
 \begin{align} \label{eq:SurplusMemberOptRef}
(\bm{d}_i^{\ast,\mbox{\tiny DNEM}},z_i^{\ast,\mbox{\tiny DNEM}})=& \underset{\bm{d}_i \in \mathbb{R}_{+}^K}{\operatorname{argmax}} ~ S_i^{\mbox{\tiny DNEM}}(\bm{d}_i,z_i):=U_i\left(\bm{d}_i\right)- \Gamma^{\mbox{\tiny DNEM}}\cdot z_i \nonumber \\
& \text { subject to } \quad 
z_i := \boldsymbol{1}^\top \bm{d}_{i}-b_i  \\
& \hspace{2cm} \underline{z}_i \leq z_i \leq \overline{z}_i \nonumber\\&\hspace{2cm} \underline{\bm{d}}_i \preceq \bm{d}_i \preceq \overline{\bm{d}}_i.\nonumber
\end{align} 
The following Lemma \ref{lem:OptSchedule} characterizes the optimal consumption and net consumption of every community member under the proposed OEs-aware mechanism.

\begin{figure}
    \centering
    \includegraphics[scale = 0.43]{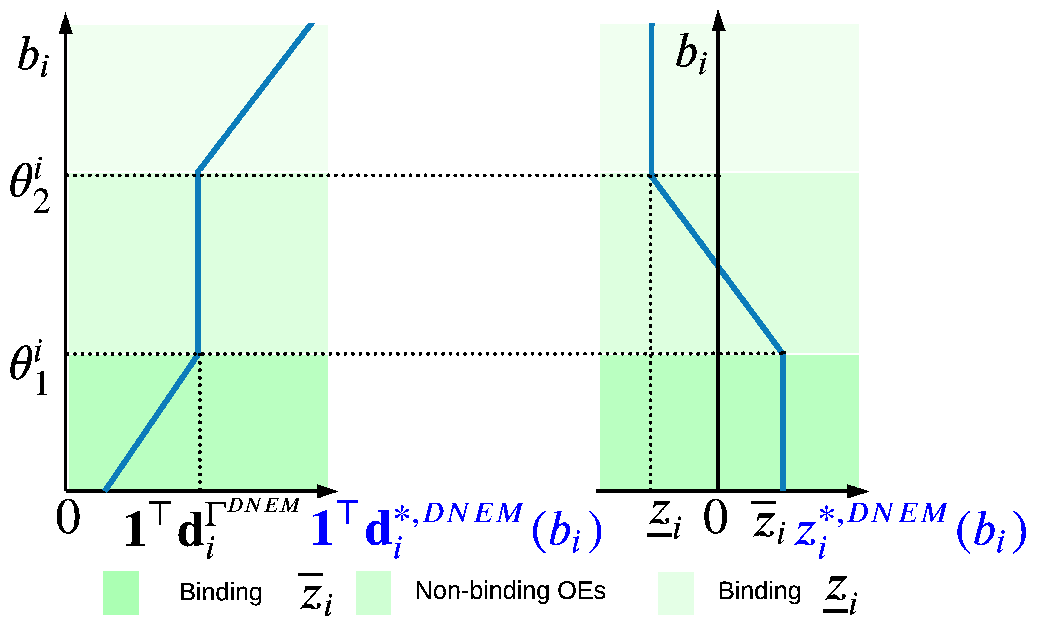}
    \caption{Community members optimal consumption and net consumption under the OEs-aware D-NEM.}
    \label{fig:OE_DNEMresponse}
\end{figure}

\begin{lemma}[Optimal member decisions]\label{lem:OptSchedule}
    Given the announced market mechanism, every member $i \in \mathcal{N}$'s optimal decisions obey a two-threshold policy with thresholds
    \begin{equation}\label{eq:MemberThresholds}
    \theta_1^i := \bm{1}^\top \bm{d}^{\Gamma^{\mbox{\tiny DNEM}}}_i - \overline{z}_i, \quad \quad \quad\theta_2^i:= \bm{1}^\top \bm{d}^{\Gamma^{\mbox{\tiny DNEM}}}_i - \underline{z}_i \geq \theta_1^i,
\end{equation}
    that schedule the consumption as
    \begin{align}\label{eq:MemberOptd}
        \bm{d}^{\ast,\mbox{\tiny DNEM}}_{i} (b_i)= \begin{cases} 
\bm{d}^{\mu_1^\ast}_i(b_i) &, b_i < \theta_1^i\\ 
\bm{d}^{\Gamma^{\mbox{\tiny DNEM}}}_i &,  b_i \in [\theta_1^i, \theta_2^i]\\
\bm{d}^{\mu_2^\ast}_i(b_i) &, b_i > \theta_2^i,
\end{cases}
\end{align}
where $\bm{d}^\Omega_{i} = \max\{\underline{\bm{d}}_i,\min\{\bm{f}_i(\bm{1} \Omega),\overline{\bm{d}}_i\} \}$ with $\Omega=\{\mu_1^\ast,\Gamma^{\mbox{\tiny DNEM}}, \mu_2^\ast\}$ and $\mu_1^\ast\geq \Gamma^{\mbox{\tiny DNEM}} \geq \mu_2^\ast$.

The prices $\mu_1^\ast(b_i)$ and $\mu_2^\ast(b_i)$ are the solutions of 
\begin{align}
    \bm{1}^\top \max\{\underline{\bm{d}}_i,\min\{\bm{f}_i(\bm{1}\mu_1),\overline{\bm{d}}_i\} \} =  \overline{z}_i +b_i, ~~~
    \bm{1}^\top\max\{\underline{\bm{d}}_i,\min\{\bm{f}_i(\bm{1}\mu_2),\overline{\bm{d}}_i\} \} = \underline{z}_i+b_i,
\end{align}
respectively, and the $\max$ and $\min$ operators are elementwise. 

The optimal net consumption for every $i\in \mathcal{N}$ is, by definition,
        \begin{align}\label{eq:MemberOptz}
        z^{\ast,\mbox{\tiny DNEM}}_i(b_i) &= \bm{1}^\top \bm{d}^{\ast,\mbox{\tiny DNEM}}_{i}(b_i) - b_i.
    \end{align}
\end{lemma}
\noindent {\em Proof:} See the appendix.\hfill$\blacksquare$\\
Lemma \ref{lem:OptSchedule} reveals how every member $i\in \mathcal{N}$ uses the announced community price $\Gamma^{\mbox{\tiny DNEM}}$ and its OEs to compute the thresholds $\theta_1^i,\theta_2^i$, which are then compared to the member's local DER $b_i$ to schedule consumption (Fig.\ref{fig:OE_DNEMresponse}). 

As depicted in Fig.\ref{fig:OE_DNEMresponse}, Lemma \ref{lem:OptSchedule}, and given the monotonicity of $\Gamma^{\mbox{\tiny DNEM}}$ and $\bm{f}_i$ for all $i\in \mathcal{N}$, shows that the optimal consumption of community members $\bm{d}^{\ast,\mbox{\tiny DNEM}}_{i}$ is monotonically increasing with $b_i$, leading to a monotonically decreasing optimal net consumption $z^{\ast,\mbox{\tiny DNEM}}_i$ with $b_i$. 

Given the optimal responses of the benchmark (Lemma 2 in the appendix) and the community member (Lemma \ref{lem:OptSchedule}), Theorem \ref{thm:IndRat} establishes {\em individual rationality} under the proposed OEs-aware D-NEM.

\begin{theorem}[Individual rationality]\label{thm:IndRat}
    Under the OEs-aware D-NEM, every member $i \in \mathcal{N}$ and for all $b_i$, achieves a surplus no less than its benchmark, i.e.,
\begin{equation}\label{eq:Sopt2Benchmark_moel2}
    S^{\ast,\mbox{\tiny DNEM}}_i(b_i) \geq S^{\ast,\mbox{\tiny NEM}}_i(b_i),
\end{equation}
where $S^{\ast,\mbox{\tiny DNEM}}_i(b_i) := S^{\ast,\mbox{\tiny DNEM}}_i(\bm{d}^{\ast,\mbox{\tiny DNEM}}_i(b_i),z^{\ast,\mbox{\tiny DNEM}}_i(b_i))$ is the member surplus under optimal decisions.
\end{theorem}
\noindent {\em Proof:} See the appendix.\hfill$\blacksquare$

Theorem \ref{thm:IndRat} shows that every member $i \in \mathcal{N}$ finds it advantageous to join the community over autonomously facing the DSO.

\section{Social optimality and Cost-Causation}\label{sec:theory}
Given the OEs-aware D-NEM and the corresponding rational prosumer response, we establish two primary results on social optimality (Theorem \ref{thm:MktEff}) and conformity with the generalized cost-causation principle (Theorem \ref{thm:equity}). 

\begin{theorem}[Decentralized welfare optimality]\label{thm:MktEff}
Under the OEs-aware D-NEM, the aggregate surplus of community members achieves the community maximum welfare, i.e., $\sum_{i\in \mathcal{N}} S^{\ast,\mbox{\tiny DNEM}}_i(b_i) = W_{\mathcal{N}}^{\dagger,\mbox{\tiny NEM}}(\bm{b})$.
\end{theorem}
\noindent {\em Proof:} See the appendix.\hfill$\blacksquare$

Theorem \ref{thm:MktEff} also shows that the maximum community welfare is wholly distributed to the members. Next, we leverage Definition \ref{def:CostCausation} to establish the conformity of the OEs-aware D-NEM that induced community members to achieve the maximum social welfare with the cost-causation principle.
\begin{theorem}[Conformity with the generalized cost-causation principle]\label{thm:equity}
    The proposed OEs-aware D-NEM satisfies the generalized cost-causation principle.
\end{theorem}
\noindent {\em Proof:} See the appendix.\hfill$\blacksquare$

Intuitively, the non-discriminatory price of the OEs-aware D-NEM directly leads to the {\em equity} and {\em monotonicity} axioms, and structuring the volumetric charge based on the member's own net consumption $z_i$ enables penalizing net-consumers ($z_i>0$), and rewarding net-producers ($z_i<0$), which satisfy axiom 5 in Definition \ref{def:CostCausation}.

\section{Numerical Study}\label{sec:num}
To evaluate the community market mechanism and the corresponding optimal prosumer response, we used a one-year DER data\footnote{We used 2018 \href{https://www.pecanstreet.org/dataport/}{PecanStreet data} for households in Austin, TX with 15-minute granularity.} of $N=20$ residential customers (3 of which do not have BTM generation) to construct an energy community, whereby the 20 households pool and aggregate their resources behind a DSO revenue meter under a NEM policy\footnote{Centralized resources were not considered.}. The DSO charges the community ({\em retail rate} $\pi^+$) based on a ToU rate with $\pi^+_{\mbox{\tiny ON}}=\$0.40$/kWh and $\pi^+_{\mbox{\tiny OFF}}=\$0.20$/kWh as on- and off-peak prices, respectively, and compensates the community ({\em export rate} $\pi^-$) based on the wholesale market price\footnote{We used the averaged 2018 real-time wholesale prices in Texas. The data is accessible at \href{https://www.ercot.com/mktinfo/prices}{ERCOT}.}. The DSO's fixed charge under NEM is assumed to be zero, i.e., $\pi^0=0$. The DSO OEs were varied, but we assumed homogeneous OEs and $\overline{z}_i = -\underline{z}_i, \forall i\in \mathcal{N}$.

For every $i \in \mathcal{N}$, the household's consumption preferences are modeled using a quadratic concave and non-decreasing utility function of the form
\begin{equation}\label{eq:UtilityForm}
   U_{ik}(d_{ik})=\left\{\begin{array}{ll}
\alpha_{ik} d_{ik}-\frac{1}{2}\beta_{ik} d_{ik}^2,\hspace{-0.2cm} &\hspace{-0.2cm} 0 \leq d_{ik} \leq \frac{\alpha_{ik}}{\beta_{ik}} \\
\frac{\alpha_{ik}^2}{2 \beta_{ik}},\hspace{-0.2cm} &\hspace{-0.2cm} d_{ik}>\frac{\alpha_{ik}}{\beta_{ik}},
\end{array} \right.
\end{equation}
for all $k \in \mathcal{K}$, where $\alpha_{ik}, \beta_{ik}$ are parameters that are learned and calibrated using historical retail prices\footnote{We used \href{https://data.austintexas.gov/stories/s/EOA-C-5-a-Austin-Energy-average-annual-system-rate/t4es-hvsj/}{Data.AustinTexas.gov} historical residential rates in Austin, TX.} and consumption\footnote{We used pre-2018 PecanStreet data for households in Austin, TX.}, and by predicating an elasticity for each load type (see appendix D in \cite{Alahmed&Tong:22EIRACM}). Two load types with two different utility functions of the form in (\ref{eq:UtilityForm}) were considered: 1) HVAC, and 2) other household loads\footnote{The elasticities of HVAC and other household loads are taken from \cite{ASADINEJAD_Elasticity:18EPSR}.}. We ignore device consumption limits $\underline{\bm{d}}_i, \overline{\bm{d}}_i, \forall i \in \mathcal{N}$.

We compared the welfare of four schemes under the same resources and number of customers. The first two involve no community (coalition), whereas the last two consider energy communities.
\begin{enumerate}
    \item {\em NEM-Benchmark}: $N$ prosumers who autonomously face the DSO's NEM, and solve (\ref{eq:BenchmarkProblem}), which yields a BTM-generation-aware consumption scheduling as shown in Lemma 2 in the appendix. The welfare of {\em NEM-Benchmark} is the aggregate maximized surplus of the $N$ prosumers.
     \item {\em NEM-Passive Benchmark}: similar to {\em NEM-Benchmark}, under this scheme, the $N$ prosumers autonomously face the DSO, but rather than optimally scheduling their resources as in Lemma 2, they use all of the BTM generation to reduce their payment (see \cite{Alahmed&Tong:22EIRACM} for a wider discussion). The welfare of {\em NEM-Passive Benchmark} is, also, the sum of the $N$ customers' surpluses.
    \item {\em D-NEM}: $N$ prosumers who form a community under the OEs-aware D-NEM. The welfare achieved under this case is as in (\ref{eq:CentralWelfare}).
    \item {\em NEM-Community}: unlike the {\em ex-ante} price set under {\em D-NEM}, the price and/or allocation rules under {\em NEM-Community} are set after the market is cleared. Therefore, the $N$ prosumers continue to consume as if they were facing the DSO's NEM\footnote{For a fairer comparison, and to improve the welfare under this case, we assume that prosumers consume similar to {\em NEM-Benchmark} rather than {\em NEM-Passive Benchmark}.}, but gain higher benefits by joining the coalition because $P^\pi(z_\mathcal{N})\leq \sum_{i\in \mathcal{N}} P^\pi(z_i)$, as shown, for example, in \cite{Chakraborty&Poolla&Varaiya:19TSG,Yang&Guoqiang&Spanos:21TSG, Han&Morstyn&McCulloch:19TPS}. The {\em NEM-Community} scheme includes allocation methods such as Shapley value \cite{Shapley&Shubik:73RAND, Azimetal:23TSG}, proportional rule \cite{Young:94GameTheoryBook}, the allocation in \cite{Chakraborty&Poolla&Varaiya:19TSG}, and the nucleolus \cite{Schotter&Schwodiauer:80JEL,Han&Morstyn&McCulloch:19TPS}, among others.
\end{enumerate}

\begin{figure*}[htbp]
    \centering
    \includegraphics[scale=0.46]{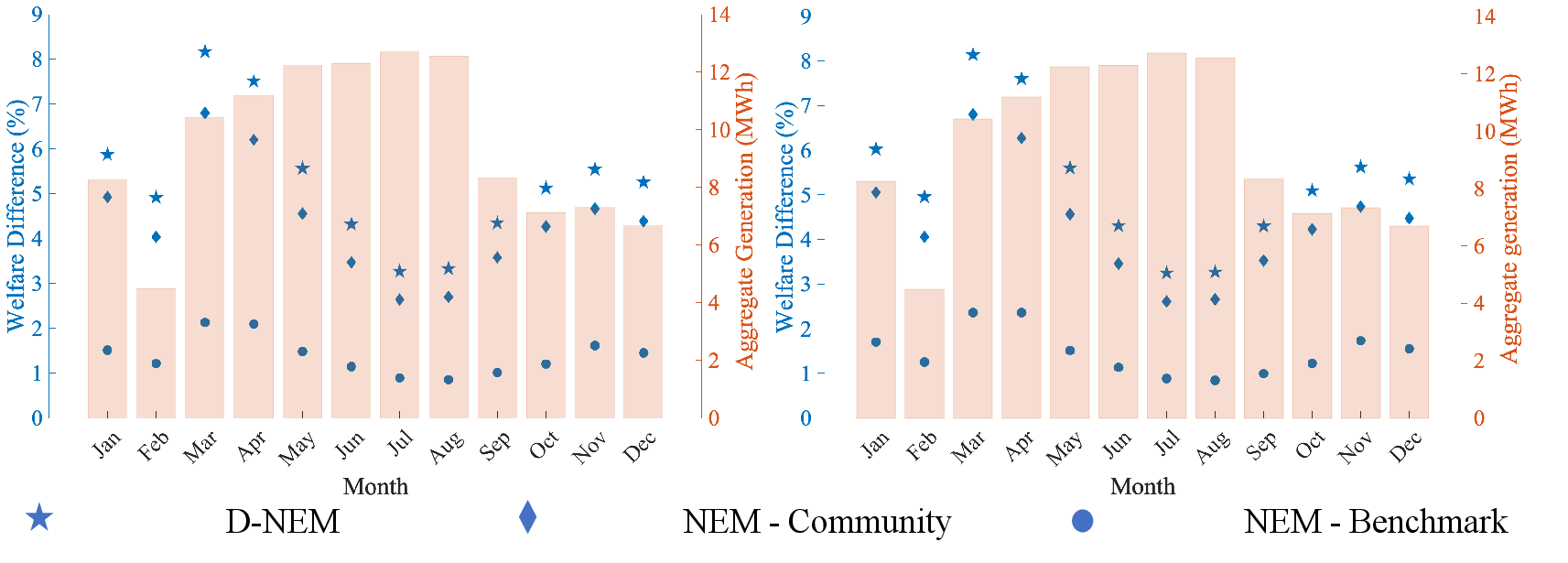}
    \caption{Monthly welfare gain (\%) over {\em NEM - Passive Benchmark}  under $-\underline{z}_i= \overline{z}_i=3$kW (left) and $-\underline{z}_i= \overline{z}_i=\infty$ (right). The bar chart shows the aggregate community generation $b_\mathcal{N}$.}
    \label{fig:WelfareZG}
\end{figure*}

The monthly welfare gains (\%) achieved by {\em NEM-Benchmark}, {\em D-NEM}, and {\em NEM-Community} over the {\em NEM-Passive Benchmark} welfare are shown in Fig.\ref{fig:WelfareZG} under $-\underline{z}_i= \overline{z}_i=3$kW OEs (left) and $-\underline{z}_i= \overline{z}_i=\infty$ (right). For every month, the difference between the circles and the reference (zero) shows the value of being an {\em active} prosumer under the DSO's NEM X, whereas the difference between the diamonds and the circles shows the value of forming the community (coalition) and sharing the resources. The asterisks show the value of adopting the OEs-aware {\em D-NEM} that induces the members to maximize global welfare.

\begin{figure}[htbp]
    \centering
    \includegraphics[scale=0.35]{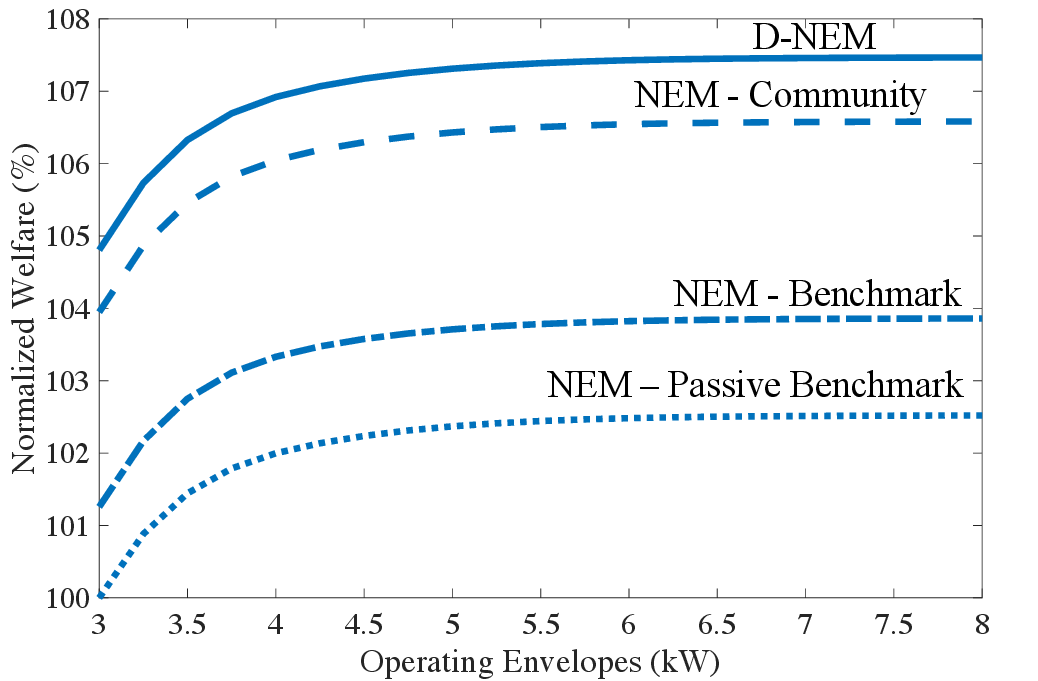}
    \caption{Normalized average monthly welfare (\%) under varying OEs.}
    \label{fig:NormWelfareOEs}
\end{figure}

Four observations in Fig.\ref{fig:WelfareZG} are in order. First, in all months, forming the energy communities achieved positive welfare gains, which shows the value of forming the coalition over autonomously facing the DSO. Second, we observe the optimality of {\em D-NEM} over the schemes, since {\em D-NEM} induces the community members to maximize global welfare. The average monthly gain under {\em D-NEM} was $\sim$5.3\%, whereas it was $\sim$1.4\% and $\sim$4.4\% under {\em NEM-Benchmark}, and {\em NEM-Community}, respectively. Third, comparing the left and right panels of Fig.\ref{fig:WelfareZG}, the welfare gains were only slightly affected by changing the OEs, because the OEs are the same regardless of whether the customer joins the community or stays under the DSO's NEM X. The OEs, however, affected the total welfare, as will be shown in Fig.\ref{fig:NormWelfareOEs}. Lastly, the welfare gains are functions of the community's aggregate renewable and also flexibility given by the utility function parameters $\alpha$ and $\beta$ in (\ref{eq:UtilityForm}). One can see that although in the summer months (June--August), renewables were the highest, hence higher welfare, the welfare gain was the lowest in those months. This is because consumption in the summer was also high, which means that the renewables were mostly consumed by BTM rather than pooled with other customers, which creates the intrinsic value of energy communities.

We observed in Fig.\ref{fig:WelfareZG} that the effect of relaxing OEs on welfare gain over {\em NEM-Passive Benchmark} was negligible. Fig. \ref{fig:NormWelfareOEs} shows the normalized average monthly welfare (to the minimum value under {\em NEM-Passive Benchmark}) of the four schemes as the OEs get relaxed from $-\underline{z}_i= \overline{z}_i=3$kW to $-\underline{z}_i= \overline{z}_i=8$kW. Under all four schemes, the welfare increased as the OEs were further relaxed. Increasing the OEs from  3kW to 8kW increased the normalized welfare of each scheme by almost 2.5\%. At $-\underline{z}_i= \overline{z}_i=3$kW, {\em D-NEM} normalized percentage welfare was $\sim$ 105\%, which increased to $\sim$ 107.5\% at $-\underline{z}_i= \overline{z}_i=8$kW, because more energy can be pooled within the community.

\section{Conclusion}\label{sec:conclusion}
This work proposes an OEs-aware market mechanism for energy communities that incorporates the DSO's dynamic OEs at each member's meter into its pricing structure to induce a collective member response that decentrally achieves the maximum global welfare while making joining the community advantageous for every member. The market mechanism charges its members by a {\em uniform}, but dynamic, price that obeys a two-threshold policy and gets announced based on how much aggregate generation-storage resources exist in the community. The OEs-aware mechanism is shown to conform with the generalized cost-causation principle of designing just and fair cost allocations. 

A potentially worthwhile future direction is to address and quantify the flexibility limitations brought by network-awareness via OEs and compare it with communities that have OEs at the PCC rather than at its members' revenue meters.

\section{Acknowledgment}
The work of Ahmed S. Alahmed and Lang Tong was supported in part by the National Science Foundation under Awards 1932501 and 2218110.
This work was authored in part by the National Renewable Energy Laboratory, operated by Alliance for Sustainable Energy, LLC, for the U.S. Department of Energy (DOE) under Contract No. DE-AC36-08GO28308. Funding is provided by the U.S. Department of Energy Office of Energy Efficiency and Renewable Energy Building Technologies Office, United States. The views expressed in the article do not necessarily represent the views of the DOE or the U.S. Government. The U.S. Government retains and the publisher, by accepting the article for publication, acknowledges that the U.S. Government retains a nonexclusive, paid-up, irrevocable, worldwide license to publish or reproduce the published form of this work, or allow others to do so, for U.S. Government purposes.

%\begin{comment}
\newpage
\section*{Appendix: Proofs and additional theoretical results}\label{sec:app}
 \subsection{Proof of Lemma \ref{lem:OptSchedule}}
The proof follows directly from KKT conditions. We drop the subscript $i$ to simplify notations. The Lagrangian of (\ref{eq:SurplusMemberOptRef}) is given by:
\begin{align*}
    \mathcal{L}(\cdot) &:= \Gamma^{\mbox{\tiny DNEM}} \cdot (\sum_{k=1}^K d_k - b) - \sum_{k=1}^K U_k(d_k) + \sum_{k=1}^K \overline{\gamma}_k (d_k - \overline{d}_k)\\& - \sum_{k=1}^K \underline{\gamma}_k (d_k - \underline{d}_k) + \overline{\lambda} (\sum_{k=1}^K d_k - b - \overline{z}) - \underline{\lambda} (\sum_{k=1}^K d_k - b - \underline{z}),
\end{align*}
where $\overline{\gamma}_k$ and $\underline{\gamma}_k$ are device $k$'s Lagrangian multipliers associated with its upper and lower consumption limits, respectively, whereas $\overline{\lambda}$ and $\underline{\lambda}$ are Lagrangian multipliers associated with the upper and lower OEs, respectively. For every device $k$, deriving the Lagrangian with respect to $d_k$, gives
$$\frac{\partial \mathcal{L}}{\partial d_k}= \Gamma^{\mbox{\tiny DNEM}} - L_k(d^{\ast,\mbox{\tiny DNEM}}_k) + \overline{\gamma}_k - \underline{\gamma}_k + \overline{\lambda} - \underline{\lambda}=0,$$
which yields
$$d^{\ast,\mbox{\tiny DNEM}}_k = f_k(\Gamma^{\mbox{\tiny DNEM}}+\overline{\gamma}_k-\underline{\gamma}_k+ \overline{\lambda} - \underline{\lambda}).$$
Given the assumption in (\ref{eq:feasibilityAssum}) and complementary slackness conditions of the consumption limits multipliers, we can re-write the equation above to
$$d^{\ast,\mbox{\tiny DNEM}}_k = \max\{\underline{d}_k,\min \{f_k(\Gamma^{\mbox{\tiny DNEM}}+ \overline{\lambda} - \underline{\lambda}),\overline{d}_k\}\}.$$
From the complementary slackness conditions of the OEs, we can write the consumption above as
\begin{equation*}
    d^{\ast,\mbox{\tiny DNEM}}_k = \begin{cases}
\max\{\underline{d}_k,\min \{f_k(\Gamma^{\mbox{\tiny DNEM}}),\overline{d}_k\}\} &\hspace{-0.35cm} \text{ if } \overline{\lambda}=\underline{\lambda} =0 \\ 
\max\{\underline{d}_k,\min \{f_k(\Gamma^{\mbox{\tiny DNEM}}+\overline{\lambda}),\overline{d}_k\}\} &\hspace{-0.35cm} \text{ if } \overline{\lambda}>0, \underline{\lambda}=0  \\ 
 \max\{\underline{d}_k,\min \{f_k(\Gamma^{\mbox{\tiny DNEM}}-\underline{\lambda}),\overline{d}_k\}\} &\hspace{-0.35cm} \text{ if } \overline{\lambda}=0, \underline{\lambda}>0.
\end{cases}
\end{equation*}
Let $d^{\Gamma^{\mbox{\tiny DNEM}}}_k:=\max\{\underline{d}_k,\min \{f_k(\Gamma^{\mbox{\tiny DNEM}}),\overline{d}_k\}\}$. Note that when $\overline{\lambda}=\underline{\lambda} =0$, we have $$b\in (\sum_{k=1}^K d^{\Gamma^{\mbox{\tiny DNEM}}}_k-\overline{z},\sum_{k=1}^K d^{\Gamma^{\mbox{\tiny DNEM}}}_k-\underline{z}).$$
If $b \leq \sum_{k=1}^K d^{\Gamma^{\mbox{\tiny DNEM}}}_k-\overline{z} \rightarrow \overline{z} \leq \sum_{k=1}^K d^{\Gamma^{\mbox{\tiny DNEM}}}_k - b$, because of the monotonicity of $f_k$, it must be that $\overline{\lambda}>0, \underline{\lambda}=0$, and we have  
\begin{equation}\label{eq:ProofLem11}
    \sum_{k=1}^K \max\{\underline{d}_k,\min \{f_k(\mu_1),\overline{d}_k\}\} - b = \overline{z}, \text{ where }~ \mu_1:=\Gamma^{\mbox{\tiny DNEM}}+\overline{\lambda},
\end{equation}
which must have a non-negative solution when $b\in [\sum_{k=1}^K \underline{d}_k-\overline{z},\sum_{k=1}^K d^{\Gamma^{\mbox{\tiny DNEM}}}_k- \overline{z}]$. Let
$$F_1(x):= \sum_{k=1}^K \max\{\underline{d}_k,\min \{f_k(x),\overline{d}_k\}\} - b - \overline{z},$$
which is a continuous and monotonically decreasing function of $b$. Since $F_1(\Gamma^{\mbox{\tiny DNEM}})\geq 0,~ F_1(\mu_{\mbox{\tiny max}})\leq 0$, where $\mu_{\mbox{\tiny max}}$ is such that
$$\sum_{k =1}^K \max\{\underline{d}_{k},\min\{f_{k}(\mu_{\mbox{\tiny max}}),\overline{d}_{k}\}\}= \sum_{k =1}^K \underline{d}_{k},$$ 
there must exist a $\mu_1^\ast \in [\Gamma^{\mbox{\tiny DNEM}},\mu_{\mbox{\tiny max}}]$ such that $F(\mu_1^\ast)=0$. The positive solution of (\ref{eq:ProofLem11}) also implies ${d}^{\mu_1^\ast}_{k}(b) \in [\underline{d}_{k},d_{k}^{\Gamma^{\mbox{\tiny DNEM}}}], \forall k$. Therefore, the optimal consumption when $b \leq \sum_{k=1}^K d^{\Gamma^{\mbox{\tiny DNEM}}}_k-\overline{z}$ is $d^{\ast,\mbox{\tiny DNEM}}_k(b)= d^{\mu_1^\ast}_k(b):= \max\{\underline{d}_k,\min \{f_k(\mu_1^\ast(b)),\overline{d}_k\}\}$.

Similarly, if $b\geq  \sum_{k=1}^K d^{\Gamma^{\mbox{\tiny DNEM}}}_k-\underline{z} \rightarrow \underline{z} \geq \sum_{k=1}^K d^{\Gamma^{\mbox{\tiny DNEM}}}_k - b$, it must be that $\overline{\lambda}=0, \underline{\lambda}>0$, and we have  
\begin{equation}\label{eq:ProofLem12}
\sum_{k=1}^K \max\{\underline{d}_k,\min \{f_k(\mu_2),\overline{d}_k\}\} - b = \underline{z},\text{ where }~ \mu_2:=\Gamma^{\mbox{\tiny DNEM}}-\underline{\lambda},
\end{equation}
which must have a non-negative solution when $b\in [\sum_{k=1}^K d^{\Gamma^{\mbox{\tiny DNEM}}}_k- \overline{z},\sum_{k=1}^K \overline{d}_k-\underline{z}]$. Let
\begin{equation}\label{eq:lem2F2}
F_2(x):= \sum_{k=1}^K \max\{\underline{d}_k,\min \{f_k(x),\overline{d}_k\}\} - b - \underline{z},
\end{equation}
which is a continuous and monotonically decreasing function of $b$. Since
$$F_2(\Gamma^{\mbox{\tiny DNEM}})\leq 0,\quad F_2(\mu_{\mbox{\tiny min}})\geq 0,$$
where $\mu_{\mbox{\tiny min}}$ is such that
$$\sum_{k =1}^K \max\{\underline{d}_{k},\min\{f_{k}(\mu_{\mbox{\tiny min}}),\overline{d}_{k}\}\}= \sum_{k =1}^K \overline{d}_{k},$$ there must exist a $\mu_2^\ast \in [\mu_{\mbox{\tiny min}},\Gamma^{\mbox{\tiny DNEM}}]$ such that $F(\mu_2^\ast)=0$. The positive solution of (\ref{eq:lem2F2}) also implies ${d}^{\mu_2^\ast}_{k}(b) \in [d_{k}^{\Gamma^{\mbox{\tiny DNEM}}},\overline{d}_{k}]$. Therefore, the optimal consumption when $b \geq  \sum_{k=1}^K d^{\Gamma^{\mbox{\tiny DNEM}}}_k-\underline{z}$ is $d^{\ast,\mbox{\tiny DNEM}}_k(b)= d^{\mu_2^\ast}_k(b):= \max\{\underline{d}_k,\min \{f_k(\mu_2^\ast(b)),\overline{d}_k\}\}$.

Letting $\theta_1:=\sum_{k=1}^K d^{\Gamma^{\mbox{\tiny DNEM}}}_k - \overline{z},~ \theta_2:= \sum_{k=1}^K d^{\Gamma^{\mbox{\tiny DNEM}}}_k - \underline{z}$, the optimal consumption can be written as
\begin{equation*}
    d^{\ast,\mbox{\tiny DNEM}}_k(b) = \begin{cases}
d^{\mu_1^\ast}_k(b) & \text{ if } b \leq \theta_1 \\ 
d^{\Gamma^{\mbox{\tiny DNEM}}}_k & \text{ if } b\in (\theta_1,\theta_2)  \\ 
 d^{\mu_2^\ast}_k(b) & \text{ if } b \geq \theta_2, \\ 
\end{cases}
\end{equation*}
and has a vector form
\begin{equation*}
    \bm{d}^{\ast,\mbox{\tiny DNEM}}(b) = \begin{cases}
\bm{d}^{\mu_1^\ast}(b) & \text{ if } b < \theta_1 \\ 
\bm{d}^{\Gamma^{\mbox{\tiny DNEM}}} & \text{ if } b\in [\theta_1,\theta_2]  \\ 
 \bm{d}^{\mu_2^\ast}(b) & \text{ if } b > \theta_2, \\ 
\end{cases}
\end{equation*}
where we used the fact that, for every $k$, $d^{\mu_1^\ast}_k(\theta_1)=d^{\Gamma^{\mbox{\tiny DNEM}}}_k$ and $d^{\mu_2^\ast}_k(\theta_2)=d^{\Gamma^{\mbox{\tiny DNEM}}}_k$.
By definition, the optimal net consumption $z^{\ast,\mbox{\tiny DNEM}}(b)$ is as shown in (\ref{eq:MemberOptz}). \hfill$\blacksquare$

\subsection{Lemma \ref{lem:BenchmarkSur} and its proof}
\begin{lemma}[Benchmark prosumer optimal response and maximum surplus]\label{lem:BenchmarkSur}
The optimal benchmark consumption of every member $i\in \mathcal{N}$ under the DSO's NEM regime abides by the following 4-thresholds,
\begin{align}
 \Delta_1^i &:= \Delta_2^i - \overline{z}_i, \quad \quad \Delta_2^i := \sum_{k=1}^K \max \{ \underline{d}_{ik},\min\{f_{ik}(\pi^+),\overline{d}_{ik}\}\}\label{eq:sigma2}\\
\Delta_3^i &:= \sum_{k=1}^K \max \{\underline{d}_{ik},\min\{f_{ik}(\pi^-),\overline{d}_{ik}\}\},\label{eq:sigma3} \quad \quad
\Delta_4^i := \Delta_3^i - \underline{z}_i.
\end{align}
as 
\begin{equation}\label{eq:OptConsStandalone}
    \bm{d}^{\ast,\mbox{\tiny NEM}}_i(b_i) :=\begin{cases}
\bm{d}^{\mu^+}_i(b_i) &, b_i\leq \Delta_1^{i}\\ 
\bm{d}^{\pi^+}_i &, b_i \in [\Delta_1^{i},\Delta_2^{i}]\\ 
\bm{d}^{\mu^o}_i(b_i) &, b_i \in [\Delta_2^{i},\Delta_3^{i}]\\  
\bm{d}^{\pi^-}_i &, b_i \in [\Delta_3^{i},\Delta_4^{i}]\\ 
\bm{d}^{\mu^-}_i(b_i) &, b_i \geq \Delta_4^{i}
\end{cases}
\end{equation}
where the static consumptions $\bm{d}^{\pi^+}_{ik}$ and $\bm{d}^{\pi^-}_{ik}$ and dynamic consumptions $\bm{d}^{\mu^+}_{ik}(b_i),\bm{d}^{\mu^o}_i(b_i),\bm{d}^{\mu^-}_i(b_i)$, for every $k \in \mathcal{K}$, are all given by, 
\begin{equation}\label{eq:d+d-}
    d^x_{ik} := \max \{ \underline{d}_{ik},\min\{f_{ik}(x),\overline{d}_{ik}\}\},
\end{equation}
 and the prices $\mu^+(b_i), \mu^o(b_i), \mu^-(b_i)$ are the solutions of:
 \begin{align}
   \sum_{k=1}^K  \max \{ \underline{d}_{ik},\min\{f_{ik}(\mu^p),\overline{d}_{ik}\}\} &= \overline{z}_i + b_i,\\
   \sum_{k=1}^K  \max \{ \underline{d}_{ik},\min\{f_{ik}(\mu^{zz}),\overline{d}_{ik}\}\} &=  b_i,\\
    \sum_{k=1}^K \max \{ \underline{d}_{ik},\min\{f_{ik}(\mu^n),\overline{d}_{ik}\}\} &= \underline{z}_i + b_i,
 \end{align}
 respectively, with the order $\mu^+(b_i)\geq \pi^+\geq \mu^o(b_i)\geq \pi^-\geq \mu^-(b_i)$.
 
 By definition, the surplus under optimal consumption is given by:
 \begin{equation}\label{eq:OptSStandalone}
    S^{\ast,\mbox{\tiny NEM}}_i(b_i) :=\begin{cases}
U_i(\bm{d}^{\mu^+}_i(b_i)) - \pi^+ \overline{z}_i &, b_i\leq \Delta_1^{i}\\ 
U_i(\bm{d}^{\pi^+}_i) - \pi^+ (\bm{1}^\top \bm{d}^{\pi^+}_i- b_i) &, b_i \in [\Delta_1^{i},\Delta_2^{i}]\\ 
U_i(\bm{d}^{\mu^o}_i(b_i)) &, b_i \in [\Delta_2^{i},\Delta_3^{i}]\\  
U_i(\bm{d}^{\pi^-}_i) - \pi^- (\bm{1}^\top \bm{d}^{\pi^-}_i- b_i) &, b_i \in [\Delta_3^{i},\Delta_4^{i}]\\ 
U_i(\bm{d}^{\mu^-}_i(b_i)) - \pi^- \underline{z}_i &, b_i \geq \Delta_4^{i}.
\end{cases}
\end{equation}

\end{lemma}

\subsubsection*{Proof of Lemma \ref{lem:BenchmarkSur}}
We drop the prosumer subscript/superscript $i$ for notational brevity. The concave and non-differentiable objective of the benchmark prosumer optimization in (\ref{eq:BenchmarkProblem}) can be divided into the following three convex programs $\mathcal{P}^{\mbox{\tiny NEM},+}, \mathcal{P}^{\mbox{\tiny NEM},-}$, and $\mathcal{P}^{\mbox{\tiny NEM},o}$, which correspond to when $z\geq 0, z\leq 0$ and $z=0$, respectively\footnote{The proof here follows the approach adopted in \cite{Alahmed&Tong:22IEEETSG}.}:
\begin{align} \label{eq:BenchmarkProblem+}
\begin{array}{lll}\mathcal{P}^{\mbox{\tiny NEM},+}: &  \underset{\bm{d} \in \mathbb{R}^K}{\rm minimize}&  \pi^+ (\bm{1}^\top \bm{d} - b)-U(\bm{d})  \\&\text{subject to} & \bm{1}^\top \bm{d} - b \geq 0 , \\&&
\underline{\bm{d}}\preceq \bm{d} \preceq \overline{\bm{d}} \\&& z \leq \overline{z}. 
\end{array}   \end{align} 
\begin{align} \label{eq:BenchmarkProblem-}
\begin{array}{lll}\mathcal{P}^{\mbox{\tiny NEM},-}: &  \underset{\bm{d} \in \mathbb{R}^K}{\rm minimize}&  \pi^- (\bm{1}^\top \bm{d} - b)-U(\bm{d})  \\&\text{subject to} & \bm{1}^\top \bm{d} - b \leq 0 , \\&&
\underline{\bm{d}}\preceq \bm{d} \preceq \overline{\bm{d}} \\&& \underline{z} \leq  z. 
\end{array}   \end{align} 
\begin{align} \label{eq:BenchmarkProblemo}
\begin{array}{lll}\mathcal{P}^{\mbox{\tiny NEM},o}: &  \underset{\bm{d} \in \mathbb{R}^K}{\rm minimize}&  -U(\bm{d})  \\&\text{subject to} & \bm{1}^\top \bm{d} - b = 0 , \\&&
\underline{\bm{d}}\preceq \bm{d} \preceq \overline{\bm{d}}. 
\end{array}   \end{align} 
Since we assumed that for every $i \in \mathcal{N}$, the OEs ($\overline{z},\underline{z}$) satisfy $\overline{z}\geq \bm{1}^\top \underline{\bm{d}} - b$ and $\underline{z}\leq \bm{1}^\top \overline{\bm{d}} - b$, an optimal solution for each of the three optimizations above exists. Because the Slater’s condition is satisfied for these optimizations, KKT conditions for optimality are necessary and sufficient. Given $b$,  the optimal schedule is the one that achieves the minimum value among $\mathcal{P}^{\mbox{\tiny NEM},+},\mathcal{P}^{\mbox{\tiny NEM},-}$ and $\mathcal{P}^{\mbox{\tiny NEM},o}$. 

\begin{enumerate}[leftmargin=*]
    \item {\em Schedule when $z\geq 0$}:\\
    Under $\mathcal{P}^{\mbox{\tiny NEM},+}$, the Lagrangian $\mathcal{L}^+$ is given by
    \begin{align*}
        \mathcal{L}^+(\cdot) &= \pi^+ (\sum_{k =1}^K d_{k} - b) - \sum_{k=1}^K U_{k}(d_{k}) + \sum_{k=1}^K \overline{\gamma}^+_{k} (d_{k}-\overline{d}_{k})- \sum_{k=1}^K\underline{\gamma}^+_{k}(d_{k}-\underline{d}_{k}) - \kappa^+ (\sum_{k=1}^K d_{k} - b)\\&+ \nu^+ (\sum_{k=1}^K d_{k} - b - \overline{z}),
    \end{align*}
    where $\overline{\gamma}^+_{k},\underline{\gamma}^+_{k},\kappa^+,\nu^+ \geq 0$ are Lagrange multipliers for the upper and lower consumption constraints for every device $k$, net-consumption zone constraint, and OE constraint, respectively. Given that $\mathcal{P}^{\mbox{\tiny NEM},o}$ covers the case when $\kappa^+>0$, we can here set $\kappa^+=0$. From the KKT conditions we have, for all $k\in \mathcal{K}$,
    \begin{equation*}
        d_{k}^{\ast,\mbox{\tiny NEM}} = f_{k}(\pi^+ + \overline{\gamma}^+_{k} - \underline{\gamma}^+_{k} + \nu^+)
    \end{equation*}
From the proof of Lemma \ref{lem:BenchmarkSur}, we know that when $\nu^+=0$ we have $$d^{\ast,\mbox{\tiny NEM}}_{k} := \max\{\underline{d}_{k},\min\{f_{k}(\pi^+),\overline{d}_{k}\}\}.$$ When the import envelope is binding, we have $\nu^+>0$, and the optimal consumption becomes $$d^{\ast,\mbox{\tiny NEM}}_{k} := \max\{\underline{d}_{k},\min\{f_{k}(\mu^p),\overline{d}_{k}\}\}, $$ where the price $\mu^p:=\pi^++\nu^+$ is such that the following equality holds:
\begin{equation}\label{eq:Lem1proofNC}
    \sum_{k=1}^K \max\{\underline{d}_{k},\min\{f_{k}(\mu^p),\overline{d}_{k}\}\}-b = \overline{z}.
\end{equation}
Next, we show that (\ref{eq:Lem1proofNC}) must have a non-negative solution when $b \in [\sum_{k=1}^K \underline{d}_k -\overline{z},\Delta_1]$. Let,
\begin{equation}\label{eq:lem1F1}
    F_1(x):= \sum_{k=1}^K\max\{\underline{d}_{k},\min\{f_{k}(x),\overline{d}_{k}\}\}-b - \overline{z}.
\end{equation}
which is a continuous and monotonically decreasing function of $b$. Since
$$F_1(\pi^+)\geq 0,\quad F_1(\mu_{\mbox{\tiny max}})\leq 0,$$
where $\mu_{\mbox{\tiny max}}$ is such that $$\sum_{k=1}^K \max\{\underline{d}_{k},\min\{f_{k}(\mu_{\mbox{\tiny max}}),\overline{d}_{k}\}\}= \sum_{k=1}^K \underline{d}_{k},$$ there must exist a $\mu^+(b) \in [\pi^+,\mu_{\mbox{\tiny max}}]$ such that $F(\mu^+(b))=0$. The positive solution of (\ref{eq:lem1F1}) also implies 
${d}^{\mu^+}_{k}(b) \in [\underline{d}_{k},d_{k}^{\pi^+}],$ and from the continuity and monotonicity of $F$ in $b$, ${d}^{\mu^+}_{k}(b)$ is also a continuous and monotonically increasing function of $b$. Hence, in summary:
\begin{align}
d^{\ast,\mbox{\tiny NEM}}_{k}(b) = \begin{cases}
d^{\pi^+}_{k}, &\hspace{-.75em} \sum_{k=1}^K d^{\pi^+}_{k} -b < \overline{z}\\ 
\max\{\underline{d}_{k},\min\{f_{k}(\mu^+(b)),\overline{d}_{k}\}\}, &\hspace{-.75em} \sum_{k=1}^K d^{\pi^+}_{k}-b \geq \overline{z}, \nonumber
\end{cases}
\end{align}
where $d^{\pi^+}_{k}:=\max\{\underline{d}_{k},\min\{f_{k}(\pi^+),\overline{d}_{k}\}\}$. From our definition of $\Delta_1$ and $\Delta_2$, we can re-write the optimal consumption above as
$$d^{\ast,\mbox{\tiny NEM}}_{k}(b) = \begin{cases}
d^{\pi^+}_{k} &, b > \Delta_1\\ 
d^{\mu^+}_k(b) &, b \leq \Delta_1,
\end{cases}$$
where $d^{\mu^+}_k(b):= \max\{\underline{d}_{k},\min\{f_{k}(\mu^+(b)),\overline{d}_{k}\}\}$.

    \item {\em Schedule when $z\leq 0$}:\\
 Under $\mathcal{P}^{\mbox{\tiny NEM},-}$, the Lagrangian $\mathcal{L}^-$ is given by
  \begin{align*}
        \mathcal{L}^- &= \pi^- (\sum_{k=1}^K d_{k} - b) - \sum_{k=1}^K U_{k}(d_{k})+ \sum_{k=1}^K \overline{\gamma}^-_{k} (d_{k}-\overline{d}_{k})- \sum_{k=1}^K \underline{\gamma}^-_{k}(d_{k}-\underline{d}_{k}) + \kappa^- (\sum_{k=1}^K d_{k} - b)\\& - \nu^- (\sum_{k=1}^K d_{k} - b - \underline{z}),
    \end{align*}
    where $\overline{\gamma}^-_{k},\underline{\gamma}^-_{k},\kappa^-,\nu^- \geq 0$ are Lagrange multipliers for the upper and lower consumption constraints for every $k$, net-consumption zone constraint, and operating envelop constraint, respectively. Following the same steps in the schedule when $z\geq 0$, we have, for all $k \in \mathcal{K}$:
    $$d^{\ast,\mbox{\tiny NEM}}_{k}(b) = \begin{cases}
d^{\pi^-}_{k}, &\hspace{-0.75em} d^{\pi^-}_{k}-b > \underline{z}\\ 
\max\{\underline{d}_{k},\min\{f_{k}(\mu^-(b)),\overline{d}_{k}\}\}, &\hspace{-0.75em} d^{\pi^-}_{k}-b \leq \underline{z}
\end{cases}$$
where $d^{\pi^-}_{k}:=\max\{\underline{d}_{k},\min\{f_{k}(\pi^-),\overline{d}_{k}\}\}$, and $\mu^-(b) \leq \pi^-$ is the price that solves:
\begin{equation*}
    \sum_{k=1}^K \max\{\underline{d}_{k},\min\{f_{k}(\mu^n),\overline{d}_{k}\}\}-b = \underline{z}. 
\end{equation*}
From the definitions of $\Delta_3$ and $\Delta_4$, we can re-write the optimal consumption above as
$$d^{\ast,\mbox{\tiny NEM}}_{k}(b) = \begin{cases}
d^{\pi^-}_{k} &, b < \Delta_4\\ 
\max\{\underline{d}_{k},\min\{f_{k}(\mu^-),\overline{d}_{k}\}\} &, b \geq \Delta_4
\end{cases}$$
Similar to $\mathcal{P}^{\mbox{\tiny NEM},+}$, we can show that there must exist $\mu^-(b) \in [\mu_{\mbox{\tiny min}},\pi^-]$ when $b \in [\Delta_4,\sum_{k=1}^K \overline{d}_k - \underline{z}]$. Also, the consumption when the export operating envelope $\underline{z}$ binds $d^{\mu^-}_{k}(b)$ is continuous and monotonically increasing in $b$, and bounded by $d^{\mu^-}_{k}(b) \in [d^{\pi^-}_{k},\overline{d}_{k}]$.

    \item {\em Schedule when $z= 0$}:\\
    Lastly, the schedule of $\mathcal{P}^o$ is the same as in \cite{Alahmed&Tong:22IEEETSG}. Therefore, the optimal consumption, when $b \in [\Delta_2, \Delta_3]$, is 
    \begin{equation*}
        d^{\ast,\mbox{\tiny NEM}}_{k}(b)= \max\{\underline{d}_{k},\min\{f_{k}(\mu^o),\overline{d}_{k}\}\}\in [\Delta_2,\Delta_3],
    \end{equation*}
    where $\Delta_2, \Delta_3$ are as defined in Lemma \ref{lem:BenchmarkSur}, and $\mu^o(b) \in [\pi^-,\pi^+]$ is the solution of:
    \begin{equation*}
    \sum_{k=1}^K \max\{\underline{d}_{k},\min\{f_{k}(\mu^{zz}),\overline{d}_{k}\}\}-b = 0. 
\end{equation*}
Combining the schedules in the three zones, we get the optimal consumption in (\ref{eq:OptConsStandalone}), which given the definition of surplus function in (\ref{eq:SurplusBench}), yields (\ref{eq:OptSStandalone}).  \hfill$\blacksquare$
\end{enumerate}

\subsection{Proof of Theorem \ref{thm:IndRat}}
The proof is done by comparing the community member surplus under optimal decisions in Lemma \ref{lem:OptSchedule}, given by

\begin{equation}\label{eq:OptSurMember}
    S^{\ast,\mbox{\tiny DNEM}}_i(b_i) = \begin{cases} 
U_i(\bm{d}^{\mu_1^\ast}_i(b_i)) - \Gamma^{\mbox{\tiny DNEM}}\cdot \overline{z}_i &, b_i < \theta_1^i\\ 
U_i(\bm{d}^{\Gamma^{\mbox{\tiny DNEM}}}_i) - \Gamma^{\mbox{\tiny DNEM}} \cdot(\bm{1}^\top \bm{d}^{\Gamma^{\mbox{\tiny DNEM}}}_i - b_i) &,  b_i \in [\theta_1^i, \theta_2^i]\\
U_i(\bm{d}^{\mu_2^\ast}_i(b_i)) - \Gamma^{\mbox{\tiny DNEM}}\cdot \underline{z}_i &, b_i > \theta_2^i,
\end{cases}
\end{equation}
to the benchmark surplus under optimal benchmark decision $S^{\ast,\mbox{\tiny NEM}}_i(b_i)$, shown in (\ref{eq:OptSStandalone}) in Lemma \ref{lem:BenchmarkSur}.

Table \ref{tab:Thm1cases} lists the 21 cases of the surplus difference between the community member and its benchmark given $b_i$ and the community price $\Gamma^{\mbox{\tiny DNEM}}(\bm{b})$, given by
$$\Delta S_i(b_i):= S^{\ast,\mbox{\tiny DNEM}}_i(b_i) - S^{\ast,\mbox{\tiny NEM}}_i(b_i). $$
The proof is complete if we show that in each case in Table \ref{tab:Thm1cases}, $\Delta S_i \geq 0$. Note that in cases 1--2, and 20--21, $\Delta S_i = 0$. Also, in cases 7, 8, 14, and 15, $\Delta S_i \geq 0$ because $\pi^+ \geq \pi^z \geq \pi^-$ and $\overline{z}_i \geq 0, \underline{z}_i\leq 0$. Note that some of the cases overlap with the cases in \cite{Alahmed&Tong:23ECjournalarXiv} where OEs are absent. Particularly, case 3 here matches piece 2 in \cite{Alahmed&Tong:23ECjournalarXiv}, case 4 matches piece 3 \cite{Alahmed&Tong:23ECjournalarXiv}, case 10 matches piece 4 \cite{Alahmed&Tong:23ECjournalarXiv}, case 11 matches piece 5 in \cite{Alahmed&Tong:23ECjournalarXiv}, case 12 matches piece 6 in \cite{Alahmed&Tong:23ECjournalarXiv}, case 18 matches piece 7 in \cite{Alahmed&Tong:23ECjournalarXiv}, and case 19 matches piece 8 in \cite{Alahmed&Tong:23ECjournalarXiv}. So it remains to prove the surplus difference non-negativity of the remaining 6 cases, which are 5--6, 9, 13, 16--17. 

\begin{table}[htbp]
\centering
\caption{Surplus difference between a community member and its benchmark. }
\label{tab:Thm1cases}
\begin{tabular}{|c|c|c|c|c|}
\hline
Case & $\Gamma^{\mbox{\tiny DNEM}}$ & Community member                      & Benchmark member                      & Surplus difference ($\Delta S_i(b_i)$) \\ \hline
1    & $\pi^+$                      & $b_i < \theta_1^i$                & $b_i<\Delta_1^i$                   &     $0$       \\ \hline
2    & $\pi^+$                      & $b_i \in [\theta_1^i,\theta_2^i]$ & $b_i \in [\Delta_1^i,\Delta_2^i]$  &     $0$      \\ \hline
3    & $\pi^+$                      & $b_i \in [\theta_1^i,\theta_2^i]$ & $b_i \in [\Delta_2^i, \Delta_3^i]$ &     $U_i(\bm{d}^{\pi^+}_i) - \pi^+ (\bm{1}^\top \bm{d}^{\pi^+}_i- b_i) - U_i(\bm{d}^{\mu^o}_i(b_i))$       \\ \hline
4    & $\pi^+$                      & $b_i \in [\theta_1^i,\theta_2^i]$ & $b_i \in [\Delta_3^i,\Delta_4^i]$  &    $U_i(\bm{d}^{\pi^+}_i) - \pi^+ (\bm{1}^\top \bm{d}^{\pi^+}_i- b_i) - U_i(\bm{d}^{\pi^-}_i) + \pi^- (\bm{1}^\top \bm{d}^{\pi^-}_i- b_i)$        \\ \hline
5    & $\pi^+$                      & $b_i > \theta_2^i$                & $b_i \in [\Delta_2^i,\Delta_3^i]$  &     $U_i(\bm{d}^{\mu_2^\ast}_i(b_i)) - \pi^+ \underline{z}_i - U_i(\bm{d}^{\mu^o}_i(b_i))$       \\ \hline
6    & $\pi^+$                      & $b_i > \theta_2^i$                & $b_i \in [\Delta_3^i,\Delta_4^i]$  &   $U_i(\bm{d}^{\mu_2^\ast}_i(b_i)) - \pi^+ \underline{z}_i - U_i(\bm{d}^{\pi^-}_i) + \pi^- (\bm{1}^\top \bm{d}^{\pi^-}_i- b_i)$         \\ \hline
7    & $\pi^+$                      & $b_i > \theta_2^i$                & $b_i > \Delta_4^i$                 &  $(\pi^- - \pi^+)\underline{z}_i $          \\ \hline
8    & $\pi^z$                      & $b_i < \theta_1^i$                & $b_i < \sigma_1^i$                 &  $(\pi^+ - \pi^z) \overline{z}_i$          \\ \hline
9    & $\pi^z$                      & $b_i < \theta_1^i$                & $b_i \in [\Delta_1^i,\Delta_2^i]$  &   $U_i(\bm{d}^{\mu_1^\ast}_i(b_i)) - \pi^z \overline{z}_i - U_i(\bm{d}^{\pi^+}_i) + \pi^+ (\bm{1}^\top \bm{d}^{\pi^+}_i- b_i)$         \\ \hline
10   & $\pi^z$                      & $b_i \in [\theta_1^i,\theta_2^i]$ & $b_i \in [\Delta_1^i,\Delta_2^i]$  &  $U_i(\bm{d}^{\pi^z}_i) - \pi^z (\bm{1}^\top \bm{d}^{\pi^z}_i - b_i) - U_i(\bm{d}^{\pi^+}_i) + \pi^+ (\bm{1}^\top \bm{d}^{\pi^+}_i- b_i)$      \\ \hline
11   & $\pi^z$                      & $b_i \in [\theta_1^i,\theta_2^i]$ & $b_i \in [\Delta_2^i,\Delta_3^i]$  &    $U_i(\bm{d}^{\pi^z}_i) - \pi^z (\bm{1}^\top \bm{d}^{\pi^z}_i - b_i) - U_i(\bm{d}^{\mu^o}_i(b_i))$        \\ \hline
12   & $\pi^z$                      & $b_i \in [\theta_1^i,\theta_2^i]$ & $b_i \in [\Delta_3^i,\Delta_4^i]$  &   $U_i(\bm{d}^{\pi^z}_i) - \pi^z (\bm{1}^\top \bm{d}^{\pi^z}_i - b_i) - U_i(\bm{d}^{\pi^-}_i) + \pi^- (\bm{1}^\top \bm{d}^{\pi^-}_i- b_i)$         \\ \hline
13   & $\pi^z$                      & $b_i > \theta_2^i$                & $b_i \in [\Delta_3^i,\Delta_4^i]$  &  $U_i(\bm{d}^{\mu_2^\ast}_i(b_i)) - \pi^z \underline{z}_i - U_i(\bm{d}^{\pi^-}_i) + \pi^- (\bm{1}^\top \bm{d}^{\pi^-}_i- b_i)$          \\ \hline
14   & $\pi^z$                      & $b_i > \theta_2^i$                & $b_i > \Delta_4^i$                 &  $(\pi^-- \pi^z) \underline{z}_i$           \\ \hline
15   & $\pi^-$                      & $b_i < \theta_1^i$                & $b_i < \Delta_1^i$                 &  $(\pi^+-\pi^-) \overline{z}_i $          \\ \hline
16   & $\pi^-$                      & $b_i < \theta_1^i$                & $b_i \in [\Delta_1^i,\Delta_2^i]$                                   &    $U_i(\bm{d}^{\mu_1^\ast}_i(b_i)) - \pi^- \overline{z}_i - U_i(\bm{d}^{\pi^+}_i) + \pi^+ (\bm{1}^\top \bm{d}^{\pi^+}_i- b_i)$        \\ \hline
17   & $\pi^-$                      & $b_i < \theta_1^i$                & $b_i \in [\Delta_2^i,\Delta_3^i]$                                   &   $U_i(\bm{d}^{\mu_1^\ast}_i(b_i)) - \pi^- \overline{z}_i - U_i(\bm{d}^{\mu^o}_i)$          \\ \hline
18   & $\pi^-$                      & $b_i \in [\theta_1^i,\theta_2^i]$ & $b_i \in [\Delta_1^i,\Delta_2^i]$                                   &    $U_i(\bm{d}^{\pi^-}_i) - \pi^- (\bm{1}^\top \bm{d}^{\pi^-}_i- b_i) - U_i(\bm{d}^{\pi^+}_i) + \pi^+ (\bm{1}^\top \bm{d}^{\pi^+}_i- b_i)$        \\ \hline
19   & $\pi^-$                      & $b_i \in [\theta_1^i,\theta_2^i]$ & $b_i \in [\Delta_2^i,\Delta_3^i]$                                   &    $U_i(\bm{d}^{\pi^-}_i) - \pi^- (\bm{1}^\top \bm{d}^{\pi^-}_i- b_i) - U_i(\bm{d}^{\mu^o}_i)$        \\ \hline
20   & $\pi^-$                      & $b_i \in [\theta_1^i,\theta_2^i]$ & $b_i \in [\Delta_3^i,\Delta_4^i]$                                   &    $0$        \\ \hline
21   & $\pi^-$                      & $b_i > \theta_2^i$                & $b_i > \Delta_4^i$                             &     $0$       \\ \hline
\end{tabular}
\end{table}

We use the following property for concavity of $U(\cdot)$ 
\beq \label{eq:ConcavityPropoerty}
L(x)\geq \frac{U(y)-U(x)}{y-x}\geq L(y).
\eeq
to show that $\Delta S_i\geq 0$, for the remaining cases.

\begin{itemize}[leftmargin=*]
    \item Case 5: Note that, from Lemmas \ref{lem:OptSchedule}-\ref{lem:BenchmarkSur} $\mu_2^\ast = \mu^- \leq \mu^o$, which given the monotonicity of $f_{ik}, \forall i,k$ means that $\bm{d}^{\mu_2^\ast}_i(b_i) \succeq \bm{d}^{\mu^o}_i(b_i)$, hence $U_i(\bm{d}^{\mu_2^\ast}_i(b_i)) \geq U_i(\bm{d}^{\mu^o}_i(b_i))$. Therefore, in case 5,
    $\Delta S_i(b_i) = U_i(\bm{d}^{\mu_2^\ast}_i(b_i)) - \pi^+ \underline{z}_i - U_i(\bm{d}^{\mu^o}_i(b_i)) \geq 0$
    because $U_i(\bm{d}^{\mu_2^\ast}_i(b_i)) - U_i(\bm{d}^{\mu^o}_i(b_i))\geq 0$ and $- \pi^+ \underline{z}_i \geq 0$.
    \item Case 6: Using the same argument in case 5, we have $U_i(\bm{d}^{\mu_2^\ast}_i(b_i)) \geq U_i(\bm{d}^{\pi^-}_i)$. The non-negativity of this piece is proved if we show that $Q :=-\pi^+ \underline{z}_i  + \pi^- (\bm{1}^\top \bm{d}^{\pi^-}_i- b_i)\geq 0$. Note that in this case, we have 
    $$\bm{1}^\top \bm{d}^{\pi^-}_i - \underline{z}_i \geq b_i \geq \bm{1}^\top \bm{d}^{\pi^+}_i - \underline{z}_i, $$
    which after subtracting $\bm{1}^\top \bm{d}^{\pi^-}_i$ and multiplying by (-1) becomes
    $$  \underline{z}_i \leq \bm{1}^\top \bm{d}^{\pi^-}_i-b_i \leq  \underline{z}_i+\bm{1}^\top \bm{d}^{\pi^-}_i-\bm{1}^\top \bm{d}^{\pi^+}_i.$$
    Using the lower bound above, we have $Q = (\pi^--\pi^+) \underline{z}_i \geq 0$. 
    \item Case 9: Here it must be that $\pi^+ \geq \mu_1^\ast$ because $b_i = \bm{1}^\top \bm{d}^{\mu^\ast_1}_i(b_i) - \overline{z}_i$ and $b_i \geq \bm{1}^\top \bm{d}^{\pi^+}_i - \overline{z}_i$, which mean that $\bm{d}^{\mu^\ast_1}_i(b_i) > \bm{d}^{\pi^+}_i$, implying $\pi^+ \geq \mu_1^\ast$ from the monotonicity of the inverse marginal utility. Using (\ref{eq:ConcavityPropoerty})  and the additivity property of the utility function ($U_i(\bm{d}_i):=\sum_{k=1}^K U_{ik}(d_{ik})$), we have
    \begin{align*}
         \sum_{k=1}^K (d^{\mu^\ast_1}_{ik}(b_{i}) - d^{\pi^+}_{ik}) L_{ik}(d^{\pi^+}_{ik}) \geq U_i(\bm{d}^{\mu^\ast_1}_i(b_i)) - U_i(\bm{d}^{\pi^+}_i) \geq \sum_{k=1}^K (d^{\mu^\ast_1}_{ik}(b_i) - d^{\pi^+}_{ik}) L_{ik}(d^{\mu^\ast_1}_{ik}(b_i)).
    \end{align*}
    By noting that $\pi^+ \geq L_{ik}(d^{\mu^\ast_1}_{ik}(b_i))=:\nu(b_i) \geq \pi^z, \forall k$, and using the lower bound above, the surplus difference becomes
    \begin{align*}
        \Delta S_i(b_i) = (\pi^+ - \nu(b_i)) \bm{1}^\top \bm{d}^{\pi^+}_i + \nu(b_i) \bm{1}^\top \bm{d}_i^{\mu^\ast_1} - \pi^z \overline{z}_i - \pi^+ b_i= (\nu(b_i) - \pi^+)(b_i - \bm{1}^\top \bm{d}^{\pi^+}_i) + (\nu(b_i) - \pi^z) \overline{z}_i \geq 0,
    \end{align*}
    where we used $\bm{1}^\top \bm{d}_i^{\mu^\ast_1} = b_i +\overline{z}_i$, and $\nu(b_i) \geq \pi^z$, and $b_i \leq \bm{1}^\top \bm{d}^{\pi^+}_i$ (from $b_i \in [\Delta_1,\Delta_2]$).
    \item Case 13: Same as in case 6, we have $U_i(\bm{d}^{\mu_2^\ast}_i(b_i)) \geq U_i(\bm{d}^{\pi^-}_i)$. The non-negativity of this piece is proved if we show that $Q :=-\pi^z \underline{z}_i  + \pi^- (\bm{1}^\top \bm{d}^{\pi^-}_i- b_i)\geq 0$. Note that in this case, we have 
    $$\bm{1}^\top \bm{d}^{\pi^-}_i - \underline{z}_i \geq b_i \geq \bm{1}^\top \bm{d}^{\pi^z}_i - \underline{z}_i, $$
    which after subtracting $\bm{1}^\top \bm{d}^{\pi^-}_i$ and multiplying by (-1) becomes
    $$  \underline{z}_i \leq \bm{1}^\top \bm{d}^{\pi^-}_i-b_i \leq  \underline{z}_i+\bm{1}^\top \bm{d}^{\pi^-}_i-\bm{1}^\top \bm{d}^{\pi^-}_i.$$
    Using the lower bound above, we have $Q = (\pi^--\pi^z) \underline{z}_i \geq 0$. 
    \item Case 16: This case is the same as 9, except $\pi^z$ is replaced with $\pi^-$. Following the same steps in case 9, the surplus difference is
    \begin{align*}
        \Delta S_i(b_i) &= (\pi^+ - \nu(b_i)) \bm{1}^\top \bm{d}^{\pi^+}_i + \nu(b_i) \bm{1}^\top \bm{d}_i^{\mu^\ast_1} - \pi^- \overline{z}_i - \pi^+ b_i\\&= (\nu(b_i) - \pi^+)(b_i - \bm{1}^\top \bm{d}^{\pi^+}_i) + (\nu(b_i) - \pi^-) \overline{z}_i \geq 0
    \end{align*}
    where  we used $\bm{1}^\top \bm{d}_i^{\mu^\ast_1} = b_i +\overline{z}_i$, and $\nu(b_i) \geq \pi^-$ from Lemma \ref{lem:OptSchedule} and $b_i \leq \bm{1}^\top \bm{d}^{\pi^+}_i$ from the condition $b_i \in [\Delta_1,\Delta_2]$.
    \item Case 17: Here we have two sub-cases
    \begin{itemize}[leftmargin=*]
        \item Case 17.1 ($\mu_1^\ast(b_i) \geq \mu^o(b_i)$): This yields $\bm{d}_i^{\mu^o}(b_i) \geq \bm{d}_i^{\mu^\ast_1}(b_i)$ and $U_i(\bm{d}_i^{\mu^o}(b_i)) \geq U_i(\bm{d}_i^{\mu^\ast_1}(b_i))$. Using (\ref{eq:ConcavityPropoerty}) and multiplying by (-1), we have
    \begin{align*}
         \sum_{k=1}^K (d^{\mu^\ast_1}_{ik}(b_i) - d^{\mu^o}_{ik}(b_i)) L_{ik}(d^{\mu^\ast_1}_{ik}(b_i)) \geq U_i(\bm{d}^{\mu^\ast_1}_i(b_i)) - U_i(\bm{d}^{\mu^o}_i(b_i)) \geq \sum_{k=1}^K (d^{\mu^\ast_1}_{ik}(b_i) - d^{\mu^o}_{ik}(b_i)) L_{ik}(d^{\mu^o}_{ik}(b_i)).
    \end{align*}
    By noting that $\pi^- \leq L_{ik}(d^{\mu^o}_{ik}(b_i))=:\nu(b_i) \leq \mu^\ast_1(b_i), \forall k$, and using the lower bound above, the surplus difference becomes
    \begin{align*}
        \Delta S_i(b_i) &= \nu(b_i) (\bm{1}^\top \bm{d}_i^{\mu^\ast_1}- \bm{1}^\top \bm{d}^{\mu^o}_i(b_i))  - \pi^- \overline{z}_i= (\nu(b_i) - \pi^-)\overline{z}_i \geq 0
    \end{align*}
     where we used $\bm{1}^\top \bm{d}_i^{\mu^\ast_1} = b_i +\overline{z}_i$ from Lemma \ref{lem:OptSchedule} and $ \bm{1}^\top \bm{d}^{\mu^o}_i=b_i$ from Lemma \ref{lem:BenchmarkSur}.
        \item Case 17.2 ($\mu_1^\ast(b_i) \leq \mu^o(b_i)$): The same method in case 17.1 is applied to show that, in this case, $\Delta S_i(b_i)\geq 0$.
    \end{itemize}
\end{itemize}
This completes the proof. \hfill$\blacksquare$

\subsection{Proof of Theorem \ref{thm:MktEff}}
The proof follows from comparing the aggregated surplus of community members under optimal decisions ($\sum_{i \in \mathcal{N}} S^{\ast,\mbox{\tiny DNEM}}_i(b_i)$) to the maximum welfare under centralized operation derived in Lemma \ref{lem:CentralizedWelfare} below. 

\begin{lemma}[Community maximum welfare]\label{lem:CentralizedWelfare}
   The maximum community welfare under centralized resource scheduling is 
\end{lemma}

\subsection*{Proof of Lemma \ref{lem:CentralizedWelfare}}

 Recall the centralized welfare maximization $\mathcal{P}^{\mbox{\tiny NEM}}_\mathcal{N}$ in (\ref{eq:CentralWelfare}), and given the forecasted renewables $\bm{b}\geq 0$,
\begin{align*}
\begin{array}{lll}\underset{\{\bm{d}_i\}_{i=1}^N,\{z_i\}_{i=1}^N}{\rm maximize}& W_\mathcal{N}^{\mbox{\tiny NEM}}(\bm{d}_1,\ldots,\bm{d}_N,\bm{z}):=\sum_{i=1}^N U_i(\bm{d}_i) - P^\pi_\mathcal{N}(z_\mathcal{N}) \\\text{subject to} & 
z_\mathcal{N}=\sum_{i=1}^N z_i= \sum_{i=1}^N (\bm{1}^\top \bm{d}_{i} -b_i) \\&\underline{\bm{d}}_i\preceq \bm{d}_i \preceq \overline{\bm{d}}_i,~~ \forall i\in \mathcal{N}\\& \underline{z}_i \leq  z_i \leq \overline{z}_i, ~~ \forall i\in \mathcal{N}\end{array}  
\end{align*} 

Given that the last constraint above and the indicator function in the objective, the optimization above can be, similar to the proof of Lemma \ref{lem:BenchmarkSur} divided into 3 convex programs:

\begin{align*}
\begin{array}{lll}\mathcal{P}_\mathcal{N}^+:\underset{\{\bm{d}_i\}_{i=1}^N,\{z_i\}_{i=1}^N}{\rm minimize}& \sum_{i=1}^N (\pi^+ (\bm{1}^\top \bm{d}_{i} -b_i) - U_i(\bm{d}_i)) \\ \hspace{0.9cm}\text{subject to} & 
z_\mathcal{N}\geq 0 \\&\underline{\bm{d}}_i\preceq \bm{d}_i \preceq \overline{\bm{d}}_i,~~ \forall i\in \mathcal{N}\\& \underline{z}_i+b_i \leq  \bm{1}^\top \bm{d}_{i} \leq \overline{z}_i+b_i, ~~ \forall i\in \mathcal{N}\end{array}   
\end{align*}
\begin{align*}
\begin{array}{lll} \mathcal{P}_\mathcal{N}^0:\underset{\{\bm{d}_i\}_{i=1}^N,\{z_i\}_{i=1}^N}{\rm minimize}& - \sum_{i=1}^N U_i(\bm{d}_i)  \\\hspace{0.9cm} \text{subject to} & 
z_\mathcal{N}= 0 \\&\underline{\bm{d}}_i\preceq \bm{d}_i \preceq \overline{\bm{d}}_i,~~ \forall i\in \mathcal{N}\\& \underline{z}_i+b_i \leq  \bm{1}^\top \bm{d}_{i} \leq \overline{z}_i+b_i, ~~ \forall i\in \mathcal{N}\end{array}   
\end{align*}
\begin{align*}
\begin{array}{lll}\mathcal{P}_\mathcal{N}^-:\underset{\{\bm{d}_i\}_{i=1}^N,\{z_i\}_{i=1}^N}{\rm minimize}& \sum_{i=1}^N (\pi^- (\bm{1}^\top \bm{d}_{i} -b_i) - U_i(\bm{d}_i)) \\ \hspace{0.9cm}\text{subject to} & 
z_\mathcal{N}\leq 0 \\&\underline{\bm{d}}_i\preceq \bm{d}_i \preceq \overline{\bm{d}}_i,~~ \forall i\in \mathcal{N}\\& \underline{z}_i+b_i \leq  \bm{1}^\top \bm{d}_{i} \leq \overline{z}_i+b_i, ~~ \forall i\in \mathcal{N}.\end{array}   
\end{align*}
The three programs are similar to the ones in the proof of Theorem 1 in \cite{Alahmed&Tong:22IEEETSG}, but with the additional dimension of $N$ members and the constraint on $\bm{1}^\top \bm{d}_i, \forall i \in \mathcal{N}$. Therefore, following the same steps in \cite{Alahmed&Tong:22IEEETSG}, it is not hard to show that the optimal centralized consumption $\bm{d}^{\dagger,\mbox{\tiny NEM}}_i$ of every community member obeys by the two thresholds
\begin{align*}
    \tilde{d}^{\pi^+}_\mathcal{N}(\bm{b}) &:=\sum_{i=1}^N \max\{\underline{z}_i+b_i,\min\{\bm{1}^\top \bm{d}^{\pi^+}_i,\overline{z}_i+b_i\}=\sigma_1\\
    \tilde{d}^{\pi^-}_\mathcal{N}(\bm{b}) &:=\sum_{i=1}^N \max\{\underline{z}_i+b_i,\min\{\bm{1}^\top \bm{d}^{\pi^-}_i,\overline{z}_i+b_i\}=\sigma_2 \geq \sigma_1,
\end{align*}
with the solution being of the form:
\begin{align}\label{eq:Lem3pieces}
 \bm{d}^{\dagger,\mbox{\tiny NEM}}_i(\bm{b}) = \begin{cases}
 \bm{y}_i^+(b_i) & ,     b_\mathcal{N}<   \tilde{d}^{\pi^+}_\mathcal{N}(\bm{b})\\ 
\bm{y}_i^z(\bm{b}) & , b_\mathcal{N}\in [\tilde{d}^{\pi^+}_\mathcal{N}(\bm{b}),\tilde{d}^{\pi^-}_\mathcal{N}(\bm{b})]\\ 
 \bm{y}_i^-(b_i) & ,     b_\mathcal{N}< \tilde{d}^{\pi^-}_\mathcal{N}(\bm{b}),
\end{cases},~~ \forall i\in \mathcal{N}
 \end{align}
 where the first, second, and third pieces in (\ref{eq:Lem3pieces}) solve $\mathcal{P}_\mathcal{N}^+, \mathcal{P}_\mathcal{N}^0$ and $\mathcal{P}_\mathcal{N}^-$, respectively, and
 \begin{align}\label{eq:Lem3pieces2}
 \sum_{i\in \mathcal{N}} \bm{1}^\top \bm{d}^{\dagger,\mbox{\tiny NEM}}_i(\bm{b}) = \begin{cases}
 \tilde{d}^{\pi^+}_\mathcal{N}(\bm{b}) & ,     b_\mathcal{N}<   \tilde{d}^{\pi^+}_\mathcal{N}(\bm{b})\\ 
b_\mathcal{N} & , b_\mathcal{N}\in [\tilde{d}^{\pi^+}_\mathcal{N}(\bm{b}),\tilde{d}^{\pi^-}_\mathcal{N}(\bm{b})]\\ 
\tilde{d}^{\pi^-}_\mathcal{N}(\bm{b}) & ,     b_\mathcal{N}< \tilde{d}^{\pi^-}_\mathcal{N}(\bm{b}).
\end{cases}
 \end{align}
 
 Note that the optimal decisions in $\mathcal{P}_\mathcal{N}^+$ and $\mathcal{P}_\mathcal{N}^-$ can be decoupled to $N$ programs of the form in (\ref{eq:SurplusMemberOptRef}), because the constraints $z_\mathcal{N}\geq 0$ and $z_\mathcal{N}\leq 0$ in $\mathcal{P}_\mathcal{N}^+$ and $\mathcal{P}_\mathcal{N}^-$, respectively, are non-binding, given that the binding case is covered in $\mathcal{P}_\mathcal{N}^0$. Therefore, when $ b_\mathcal{N}<   \tilde{d}^{\pi^+}_\mathcal{N}(\bm{b})$ or $b_\mathcal{N}> \tilde{d}^{\pi^-}_\mathcal{N}(\bm{b})$, the operator schedules every member $i$ same as that in Lemma \ref{lem:OptSchedule}, but with $\pi^+$ or $\pi^-$, respectively, instead of $\Gamma^{\mbox{\tiny DNEM}}$. 

In the net-zero zone, i.e., $b_\mathcal{N}\in [\tilde{d}^{\pi^+}_\mathcal{N}(\bm{b}),\tilde{d}^{\pi^-}_\mathcal{N}(\bm{b})]$, the schedule of every member is also same as that in Lemma \ref{lem:OptSchedule}, but with replacing $\Gamma^{\mbox{\tiny DNEM}}$ by the price $\pi^{z\dagger}\in (\pi^-,\pi^+)$ that solves the following
 \begin{align*}
     \sum_{i=1}^N \max \left\{\underline{x}_i, \min \left\{\bm{1}^\top \max \{\bm{d}_i,\min\{\bm{f}_i(\bm{1}\mu),\overline{d}_i \}, \overline{x}_i\right\}\right\}=b_{\mathcal{N}},
 \end{align*}
 where $\underline{x}_i:=\underline{z}_i+b_i$ and $\overline{x}_i:= \overline{z}_i+b_i$. One can see that $\pi^{z\dagger}(\bm{b})$ is equal to $\pi^z(\bm{b})$ of DNEM.

 Therefore, in summary, the optimal consumption under centralized control is
 \begin{align*}
 \bm{d}^{\dagger,\mbox{\tiny NEM}}_i(\bm{b}) = \begin{cases}
 \bm{d}_i^{\dagger \pi^+}(b_i) & ,     b_\mathcal{N}<   \tilde{d}^{\pi^+}_\mathcal{N}(\bm{b})\\ 
\bm{d}_i^{\dagger \pi^z}(b_i) & , b_\mathcal{N}\in [\tilde{d}^{\pi^+}_\mathcal{N}(\bm{b}),\tilde{d}^{\pi^-}_\mathcal{N}(\bm{b})]\\ 
 \bm{d}_i^{\dagger \pi^-}(b_i) & ,     b_\mathcal{N}< \tilde{d}^{\pi^-}_\mathcal{N}(\bm{b})
\end{cases},~~ \forall i\in \mathcal{N}
 \end{align*}
where for ever $\bm{d}_i^{\dagger \pi^+}(b_i), \bm{d}_i^{\dagger \pi^z}(b_i)$, and $\bm{d}_i^{\dagger \pi^-}(b_i)$ are computed as in Lemma \ref{lem:OptSchedule}, but with replacing $\Gamma^{\mbox{\tiny DNEM}}$ by $\pi^+, \pi^z$, and $\pi^-$, respectively.

By definition, the welfare under optimal consumption decisions is
 \begin{align}\label{eq:WelfareLem3}
 W^{\dagger,\mbox{\tiny NEM}}(\bm{b}) = \begin{cases}
 \sum_{i=1}^N U_i(\bm{d}_i^{\dagger \pi^+}(b_i)) - \pi^+\cdot (\tilde{d}^{\pi^+}_\mathcal{N}(\bm{b}) - b_\mathcal{N}) &  ,    b_\mathcal{N}<   \tilde{d}^{\pi^+}_\mathcal{N}(\bm{b})\\ 
\sum_{i=1}^N U_i(\bm{d}_i^{\dagger \pi^z}(b_i)) & , b_\mathcal{N}\in [\tilde{d}^{\pi^+}_\mathcal{N}(\bm{b}),\tilde{d}^{\pi^-}_\mathcal{N}(\bm{b})]\\ 
 \sum_{i=1}^N U_i(\bm{d}_i^{\dagger \pi^-}) - \pi^- \cdot(\tilde{d}^{\pi^-}_\mathcal{N}(\bm{b}) - b_\mathcal{N}) & ,     b_\mathcal{N}< \tilde{d}^{\pi^-}_\mathcal{N}(\bm{b}).
\end{cases}
 \end{align}
\hfill$\blacksquare$

Now it remains to compare the welfare under optimal decisions in (\ref{eq:WelfareLem3}) in the proof of Lemma \ref{lem:CentralizedWelfare}, to the aggregate surplus of community members under the OE-aware D-NEM (in (\ref{eq:OptSurMember}))
\begin{equation*}
    S^{\ast,\mbox{\tiny DNEM}}_i(b_i) = \begin{cases} 
U_i(\bm{d}^{\mu_1^\ast}_i(b_i)) - \Gamma^{\mbox{\tiny DNEM}}\cdot \overline{z}_i &, b_i < \theta_1^i\\ 
U_i(\bm{d}^{\Gamma^{\mbox{\tiny DNEM}}}_i) - \Gamma^{\mbox{\tiny DNEM}}\cdot (\bm{1}^\top \bm{d}^{\Gamma^{\mbox{\tiny DNEM}}}_i - b_i) &,  b_i \in [\theta_1^i, \theta_2^i]\\
U_i(\bm{d}^{\mu_2^\ast}_i(b_i)) - \Gamma^{\mbox{\tiny DNEM}}\cdot \underline{z}_i &, b_i > \theta_2^i,
\end{cases}
\end{equation*}
which we can succinctly write as
$$S^{\ast,\mbox{\tiny DNEM}}_i(b_i) = U(\bm{d}^{\ast,\mbox{\tiny DNEM}}_{i} (b_i)) - \Gamma^{\mbox{\tiny DNEM}} \cdot (\bm{1}^\top \bm{d}^{\ast,\mbox{\tiny DNEM}}_{i} (b_i)- b_i),$$
where $\bm{d}^{\ast,\mbox{\tiny DNEM}}_{i} (b_i)$ is as in (\ref{eq:MemberOptd}) in Lemma \ref{lem:OptSchedule}. When $\Gamma^{\mbox{\tiny DNEM}}=\pi^+$, we have $\bm{d}^{\ast,\mbox{\tiny DNEM}}_i(b_i)=\bm{d}^{\dagger \pi^+}_i(b_i)$, therefore
$$S^{\ast,\mbox{\tiny DNEM}}_i(b_i) = U_i(\bm{d}^{\dagger \pi^+}_i(b_i)) - \pi^+ \cdot (\bm{1}^\top \bm{d}^{\dagger \pi^+}_i(b_i)- b_i)$$
and summing over $N$ gives
\begin{equation}\label{eq:Thm2SumN}
    \sum_{i=1}^N S^{\ast,\mbox{\tiny DNEM}}_i(b_i) = \sum_{i=1}^N U_i(\bm{d}^{\dagger \pi^+}_i(b_i)) - \pi^+ \cdot (\tilde{d}^{\pi^+}_{\mathcal{N}}- b_\mathcal{N}),
\end{equation}
where we used $\sum_{i\in \mathcal{N}} \bm{1}^\top \bm{d}^{\dagger \pi^+}_i(b_i)=\tilde{d}^{\pi^+}_{\mathcal{N}}$ from (\ref{eq:Lem3pieces2}) in Lemma \ref{lem:CentralizedWelfare}. Since the price is $\pi^+$ it must be that $b_\mathcal{N}<\sigma_1 = \tilde{d}^{\pi^+}_{\mathcal{N}}$, under which 
$W^{\dagger,\mbox{\tiny NEM}}(\bm{b})$ becomes
\begin{equation}\label{eq:Thm2Wlast}
W^{\dagger,\mbox{\tiny NEM}}(\bm{b}) = \sum_{i=1}^N U_i(\bm{d}_i^{\dagger \pi^+}(b_i)) - \pi^+ (\tilde{d}^{\pi^+}_\mathcal{N}(\bm{b}) - b_\mathcal{N})=  \sum_{i=1}^N S^{\ast,\mbox{\tiny DNEM}}_i(b_i).
\end{equation}
The same steps are followed to prove the cases when $\Gamma^{\mbox{\tiny DNEM}}=\pi^z(\bm{b})$ and $\Gamma^{\mbox{\tiny DNEM}}=\pi^-$. \hfill $\blacksquare$

\subsection{Proof of Theorem \ref{thm:equity}}
To prove that the market mechanism conforms with the cost-causation principle, we need to show that the five cost-causation axioms are satisfied. 

\begin{enumerate}[leftmargin=*]
\item {\em Individual rationality}: The market mechanism achieves individual rationality as shown in Theorem \ref{thm:IndRat}.
\item {\em Profit-neutrality}: To prove the operator's profit-neutrality under the OEs-aware D-NEM, we need to show that for all $b_i$,
$$\sum_{i=1}^N P^{\mbox{\tiny DNEM}}_i (z^{\ast,\mbox{\tiny DNEM}}_i(b_i)) = P^{\mbox{\tiny NEM}}_\mathcal{N}(\sum_{i=1}^N  z^{\ast,\mbox{\tiny DNEM}}_i(b_i)).$$

The aggregate payment of community members under the OEs-aware market mechanism is
\begin{align*}
\sum_{i=1}^N P^{\mbox{\tiny DNEM}}_i (&z^{\ast,\mbox{\tiny DNEM}}_i(b_i)) = \Gamma^{\mbox{\tiny DNEM}} \cdot \sum_{i=1}^N  z^{\ast,\mbox{\tiny DNEM}}_i(b_i)= \begin{cases}
 \pi^+ \cdot \sum_{i=1}^N (\bm{1}^\top \bm{d}_i^{\dagger \pi^+}(b_i) - b_i) & ,     b_\mathcal{N}< \sigma_1(\bm{b})  \\ 
\pi^z (\bm{b})\cdot  \sum_{i=1}^N (\bm{1}^\top \bm{d}_i^{\dagger \pi^z}(b_i) - b_i) & , b_\mathcal{N}\in [\sigma_1(\bm{b}),\sigma_2(\bm{b})]\\ 
\pi^-  \cdot \sum_{i=1}^N (\bm{1}^\top \bm{d}_i^{\dagger \pi^-}(b_i) - b_i) & , b_\mathcal{N} > \sigma_2(\bm{b}),
\end{cases}\nonumber
\end{align*}
which, given 
$$\sum_{i\in \mathcal{N}} \bm{1}^\top \bm{d}^{\dagger \pi^+}_i(b_i)=\tilde{d}^{\pi^+}_{\mathcal{N}},~~ \sum_{i\in \mathcal{N}} \bm{1}^\top \bm{d}^{\dagger \pi^z}_i(b_i)=b_{\mathcal{N}},~~ \sum_{i\in \mathcal{N}} \bm{1}^\top \bm{d}^{\dagger \pi^-}_i(b_i)=\tilde{d}^{\pi^-}_{\mathcal{N}}$$
from (\ref{eq:Lem3pieces2}) in Lemma \ref{lem:CentralizedWelfare}, and that $\sigma_1(\bm{b})=\tilde{d}^{\pi^+}_\mathcal{N}(\bm{b}), \sigma_2(\bm{b})=\tilde{d}^{\pi^-}_\mathcal{N}(\bm{b})$, can be reformulated to
$$\sum_{i=1}^N P^{\mbox{\tiny DNEM}}_i (z^{\ast,\mbox{\tiny DNEM}}_i(b_i))=\begin{cases}
 \pi^+  (\tilde{d}^{\pi^+}_{\mathcal{N}}(\bm{b}) - b_\mathcal{N}) & ,     b_\mathcal{N}< \tilde{d}^{\pi^+}_\mathcal{N}(\bm{b})  \\ 
0 & , b_\mathcal{N}\in [\tilde{d}^{\pi^+}_\mathcal{N}(\bm{b}),\tilde{d}^{\pi^-}_\mathcal{N}(\bm{b})]\\ 
\pi^-  (\tilde{d}^{\pi^-}_{\mathcal{N}}(\bm{b}) - b_\mathcal{N}) & , b_\mathcal{N} > \tilde{d}^{\pi^+}_\mathcal{N}(\bm{b}).
\end{cases}$$
The payment from the operator to the DSO under NEM X has the form in (\ref{eq:Pcommunity}), which under the consumption of community members becomes
\begin{align*}
P^{\mbox{\tiny NEM}}_\mathcal{N}(\sum_{i=1}^N  z^{\ast,\mbox{\tiny DNEM}}_i(b_i)) &= \Gamma^{\ast,\mbox{\tiny NEM}} \sum_{i=1}^N  z^{\ast,\mbox{\tiny DNEM}}_i(b_i)= \begin{cases}
\pi^+ \tilde{d}^{\pi^+}_{\mathcal{N}}(\bm{b}) - b_\mathcal{N} & ,     b_\mathcal{N}< \tilde{d}^{\pi^+}_\mathcal{N}(\bm{b})  \\ 
0 &\hspace{-0.2cm} , b_\mathcal{N}\in [\tilde{d}^{\pi^+}_\mathcal{N}(\bm{b}),\tilde{d}^{\pi^-}_\mathcal{N}(\bm{b})]\\ 
\pi^- \tilde{d}^{\pi^-}_{\mathcal{N}}(\bm{b}) - b_\mathcal{N} & , b_\mathcal{N} > \tilde{d}^{\pi^+}_\mathcal{N}(\bm{b}).
\end{cases}\\& = \sum_{i=1}^N P^{\mbox{\tiny DNEM}}_i (z^{\ast,\mbox{\tiny DNEM}}_i(b_i)).
\end{align*}

\item {\em Equal treatment of equals}: Under the OEs-aware D-NEM, the payment of any two members $i,j$ is, respectively,
$$P^{\mbox{\tiny DNEM}}_i(z^{\ast,\mbox{\tiny DNEM}}_i(b_i)) = \Gamma^{\mbox{\tiny DNEM}} \cdot z^{\ast,\mbox{\tiny DNEM}}_i(b_i),~~~P^{\mbox{\tiny DNEM}}_j(z^{\ast,\mbox{\tiny DNEM}}_j(b_j)) = \Gamma^{\mbox{\tiny DNEM}} \cdot z^{\ast,\mbox{\tiny DNEM}}_j(b_j).$$ 
Given that the DNEM price $\Gamma^{\mbox{\tiny DNEM}}$ is uniform, having $z^{\ast,\mbox{\tiny DNEM}}_i(b_i) = z^{\ast,\mbox{\tiny DNEM}}_j(b_j)$, for all $b_i$, yields $$P^{\mbox{\tiny DNEM}}_i(z^{\ast,\mbox{\tiny DNEM}}_i(b_i))=\Gamma^{\mbox{\tiny DNEM}} \cdot z^{\ast,\mbox{\tiny DNEM}}_i(b_i) = \Gamma^{\mbox{\tiny DNEM}} \cdot z^{\ast,\mbox{\tiny DNEM}}_j(b_j) = P^{\mbox{\tiny DNEM}}_j(z^{\ast,\mbox{\tiny DNEM}}_j(b_j)).$$

\item {\em Monotonicity}: Under the OEs-aware D-NEM, the payment of any two members $i,j$ is, respectively,
$$|P^{\mbox{\tiny DNEM}}_i(z^{\ast,\mbox{\tiny DNEM}}_i(b_i))| = \Gamma^{\mbox{\tiny DNEM}} \cdot |z^{\ast,\mbox{\tiny DNEM}}_i(b_i)|,~~~|P^{\mbox{\tiny DNEM}}_j(z^{\ast,\mbox{\tiny DNEM}}_j(b_j))| = \Gamma^{\mbox{\tiny DNEM}} \cdot |z^{\ast,\mbox{\tiny DNEM}}_j(b_j)|.$$ 

Note that if $|z^{\ast,\mbox{\tiny DNEM}}_i(b_i)| \geq |z^{\ast,\mbox{\tiny DNEM}}_j(b_j)|$ and $z^{\ast,\mbox{\tiny DNEM}}_j(b_j) \cdot z^{\ast,\mbox{\tiny DNEM}}_j(b_j)\geq 0$, we have
$$|P^{\mbox{\tiny DNEM}}_i(z^{\ast,\mbox{\tiny DNEM}}_i(b_i))| = \Gamma^{\mbox{\tiny DNEM}} \cdot |z^{\ast,\mbox{\tiny DNEM}}_i(b_i)|\geq \Gamma^{\mbox{\tiny DNEM}} \cdot |z^{\ast,\mbox{\tiny DNEM}}_j(b_j)|=|P^{\mbox{\tiny DNEM}}_j(z^{\ast,\mbox{\tiny DNEM}}_j(b_j))|.$$ 

\item {\em Cost causation penalty and cost mitigation reward}: Let $z^{\ast,\mbox{\tiny DNEM}}_i(b_i)> 0$, then we have $P^{\mbox{\tiny DNEM}}_i(z^{\ast,\mbox{\tiny DNEM}}_i(b_i))=\Gamma^{\mbox{\tiny DNEM}} \cdot z^{\ast,\mbox{\tiny DNEM}}_i(b_i) > 0$, because $\Gamma^{\mbox{\tiny DNEM}}>0$ . Similarly, if $z^{\ast,\mbox{\tiny DNEM}}_i(b_i)< 0$, then we have $P^{\mbox{\tiny DNEM}}_i(z^{\ast,\mbox{\tiny DNEM}}_i(b_i))=\Gamma^{\mbox{\tiny DNEM}} \cdot z^{\ast,\mbox{\tiny DNEM}}_i(b_i) < 0$.
\end{enumerate}
From Definition \ref{def:CostCausation}, given that the market mechanism satisfies all five axioms, it is that the market mechanism conforms with the cost-causation principle. \hfill$\blacksquare$
%\end{comment}

{
\bibliographystyle{IEEEtran}
\bibliography{CS_PESGM}
}

\end{document}